\documentclass[11pt,a4paper,onecolumn]{article}
\usepackage[utf8]{inputenc} 
\usepackage{amssymb,amsmath,amsthm,graphicx}
\theoremstyle{plain}
\usepackage{epstopdf}
\usepackage{hyperref} 
\usepackage{natbib}
\usepackage{enumitem} 
\usepackage{booktabs} 
\usepackage{mathtools} 
\usepackage{pdflscape} 
\usepackage{breqn} 
\usepackage{microtype}  
\usepackage{authblk}
\usepackage{a4wide}
\usepackage{multirow}
\usepackage{placeins}
\usepackage[table]{xcolor}
\usepackage{subfig}

\newtheorem{theorem}{Theorem}

\newtheorem*{definition*}{Definition}
\newtheorem{lemma}{Lemma}
\newcommand{\IF}{\mbox{IF}}
\newcommand{\FRB}{\mbox{FRB }}
\newcommand{\CDR}{\mbox{CDR }}
\newcommand{\var}{\mbox{Var}}
\newcommand{\Prob}[1]{\mbox{P}\left( #1 \right)}
\newcommand{\thetab}{\boldsymbol{\theta}}

\DeclarePairedDelimiter{\floor}{\lfloor}{\rfloor}

\begin{document}

\title{Robust bootstrap procedures for the chain-ladder method}
\author[1]{Kris Peremans}
\author[1]{Pieter Segaert }
\author[1]{Stefan Van Aelst}
\author[1]{Tim Verdonck\thanks{corresponding author. E-mails : kris.peremans@kuleuven.be, pieter.segaert@kuleuven.be, stefan.vanaelst@kuleuven.be, tim.verdonck@kuleuven.be}}
\affil[1]{Department of Mathematics, KU Leuven, Leuven, Belgium\thanks{KU Leuven, Department of Mathematics, Celestijnenlaan 200B box 2400, BE-3001 Leuven}}

\date{}

\maketitle

\begin{abstract}

Insurers are faced with the challenge of estimating the future reserves 
needed to handle historic and outstanding claims that are not fully settled. A 
well-known and widely used technique is the chain-ladder method, which is
a deterministic algorithm. To include a stochastic component one may apply 
generalized linear models to the run-off triangles based on past claims data. 
Analytical expressions for 
the standard deviation of the resulting reserve estimates are typically 
difficult to derive.  A popular alternative approach to obtain inference  
is to use the bootstrap technique. However, the standard procedures are very 
sensitive to the possible presence of outliers. These atypical observations, 
deviating from the pattern of the majority of the data, may both inflate or 
deflate traditional reserve estimates and corresponding inference such as their standard errors. 
Even when paired with a robust chain-ladder method, classical bootstrap inference may break down. Therefore, we discuss and implement several robust 
bootstrap procedures in the claims reserving framework and we investigate and 
compare their performance on both simulated and real data. We also illustrate 
their use for obtaining the distribution of one year risk measures. 

\end{abstract}

\textbf{Keywords:} Claims reserving, Generalized linear models, Outliers, Inference, Solvency II

\newpage


\section{Introduction} 
\label{sec:Introduction}

One of the main challenges for insurers is the task of estimating the 
future reserves needed to handle the liabilities related to current insurance 
contracts.  The recent Solvency II directive provides new guidelines 
for the calculation of these provisions. A comprehensive discussion on the 
Solvency II directive and its implications may be found in \cite{BAJ:SolvencyII}. They report that technical provisions under Solvency II consist of the following three components: claims provisions, premium provisions 
and a risk margin. Whereas claims provisions relate to claim events that have 
already occurred, the premium provisions relate to all future claim events 
covered by current insurance and reinsurance obligations. The risk margin is 
described as the amount needed to make the value of the technical provisions 
equivalent to the amount a (re)insurance undertaking would be 
expected to require in order to take over and meet (re)insurance 
obligations.

In this article we focus on the claims provision component. Although micro-level claims reserving models (see for example \cite{
Norberg:prediction,Antonio:micro}) seem to be very promising, \cite{ BAJ:SolvencyII} 
state that the current norm is to use deterministic actuarial techniques to 
asses the undiscounted element of the earned claim reserves and that this is 
likely to continue in the immediate future. The most popular and widely used 
technique is the chain-ladder method. A substantial amount of research has 
already been devoted to exploring the relationship between the chain-ladder method and various stochastic models (see for example  \cite{
Mack:DistFreeChainladder, Mack:VariabilityChainladder}, \cite{
Renshaw:StochModel}, \cite{ Verrall:EstLogLin, Verrall:StochClaims} and \cite{
Schiegl:ApplChainladder}). 

To approach claims reserving in a stochastic manner, the chain-ladder method is typically reformulated within the context of generalized 
linear models (GLM). However, assessing the variability of claims reserving predictions and the construction of upper limits with a certain confidence level still remains a difficult task. 
Analytical expressions for standard errors in GLM depend on strong model assumptions   (\cite{ Mack:MSEchainladder, Venter:DiscussionMack} and \cite{Wutrich:BoundsEstimationError}). 
An alternative approach to statistical inference that requires less assumptions is the bootstrap technique (see for example \cite{
Efron:BootStandError}, \cite{ Hall:BlockBootstrap} and \cite{
Chatterjee:BootstrapLasso}). Also in the claims reserving framework, the 
bootstrap technique is very popular and has been studied by various authors 
such as \cite{ Ashe:Essay}, \cite{ 
Lowe:PracticalGuide}, \cite{Pinheiro:BootstrapClaims}, \cite{ 
England:PredictDistLia} and \cite{Barnett:TrendFamily}.

When applying a claims reserving method in practice it may occur that one or a couple of observations deviate from the main pattern. One might naively expect that if the model assumptions hold for almost all available data, then the results of the classical method also hold approximately. This is unfortunately not the case and it is well known in the robust statistics literature that outlier(s) may heavily influence classical statistical techniques. \cite{Verdonck:IFClaims} have shown theoretically that the classical chain-ladder method is very sensitive to outliers and several robust methods have already been developed in the claims reserving framework (see e.g.\ \cite{Brazauskas:severity}, \cite{Brazauskas:loss}, \cite{Verdonck:RobChainladder}, 
\cite{Verdonck:IFClaims}, \cite{Verdonck:bivariate} and \cite{Pitselis:Robust}). Robust methods 
provide estimates for the claim provisions which resemble the classical estimates that would have been obtained if there were no outliers in the data, while they do not model the outlier generating process. As a consequence of fitting the majority of the data well, robust methods also provide a reliable method to detect outliers. Observations which are  flagged as outliers can then be examined in detail by experts to understand their origin. If these flagged observations are errors or one-time events, then the robust method indeed yields reliable estimates of the  claim provisions as desired.  If the flagged observations are not errors and such deviations may reoccur in the future, then the robust reserve may need to be adjusted to cover such potentially large deviations in the future. We believe that expertise from practitioners in the field of the data is needed to achieve this, possibly combined with techniques from extreme value theory. The idea of combining robust statistics (which typically downweights atypical points) and extreme value statistics (which models the extremes) has also been advocated in \citet{DellAquila:ExtremesRobust} and \citet{Hubert:Pareto}. It is however outside the scope of this paper to investigate estimators for the extra amount that needs to be added to the robust reserve estimate in such case.

To our knowledge, robust methods in claims reserving have only focused on reliable point estimates and/or outlier detection. However, confidence intervals may also become very unreliable (incorrect coverage)  and/or uninformative (very large interval length) when the data are contaminated.
In this article we investigate the influence of outliers on the variability of point estimators and we present some state-of-the-art robust bootstrap procedures in the claims reserving framework. These bootstrap techniques provide a complete approximate predictive distribution, from which characteristics such as the mean, variance and quantiles are easily derived. Note that we do not advocate replacing the classical methodology by the robust counterpart, but we advise to always apply both techniques. If the classical and robust method give approximately the same results (and no outliers are detected by the robust method), then it seems safe to continue the analysis with the traditional reserving techniques that will give the most efficient estimates. However, if both methods yield different results, then the robust procedure is helpful to gain insight into the data and will yield more reliable reserve estimates if the outliers are indeed isolated events.

This paper is structured as follows. Section \ref{sec:ChainLadderGLMBootstrapping} reviews the chain-ladder method and its formulation as a Poisson GLM (which is the default approach to obtain prediction under Solvency II). In this section, we also present the bootstrap procedure in the context of claims reserving. The robust chain-ladder method is briefly described in Section \ref{sec:RobchainladderBootstrapping} and it is shown that the classical bootstrap method may produce unreliable results when the data contains outlier(s), even when the classical chain-ladder method is replaced by its robust alternative. Therefore, we present several state-of-the-art robust bootstrap methods for GLM within the context of claims reserving in section \ref{sec:RobustBootstrapMethods}. Their performance is compared by means of a simulation study in section \ref{sec:SimulationStudy} and on a real dataset from a non-life business line in section \ref{sec:RealDataApplication}. We also illustrate how the methodology can be applied to investigate one year risk measures and to detect outliers automatically. Some concluding remarks and potential directions for further research are given in section \ref{sec:Conclusions}.

\section{Classical chain-ladder method and bootstrapping}
\label{sec:ChainLadderGLMBootstrapping}


In this section we introduce the classical chain-ladder method and briefly explain how this method can be derived from a basic GLM. Then, we describe the classical bootstrap procedure to obtain inference corresponding to the estimates in this GLM. Finally, we briefly describe the robust GLM estimator. 

\subsection{Classical chain-ladder method}
\label{sec:ChainLadderMethodAsAGLM}

Insurers have to build up reserves enabling them to pay outstanding claims 
and to meet claims arising in the future on the written contracts. The 
chain-ladder method is a very popular algorithm to compute these reserve 
estimates.  Let $Y _{ij}$ and $ C_{ij}$ denote respectively the incremental 
and cumulative claim amount corresponding to accident year $i$ and 
development year $j$ for $1 \leqslant i \leqslant I $ and $1 \leqslant j 
\leqslant J $. We suppose that $I=J=n$ for ease of notation, but all 
methodology remains valid when $I\neq J$.   Variables for which $i+ j \leqslant n +1$ correspond to past claims data and are used to predict the future claims which correspond to the variables with $i+j > n+1$. We define the total or overall reserve $R$ as the sum over all future (incremental) claims\footnote{Note that we define the total reserve $R$ as the total amount of the incremental claims in the lower right triangle (as in key references \cite{England:StochClaimsInsurance} and \cite{Wuthrich:StochasticClaims}), but in practice the final reserve is an amount set by the insurer based on this value.}:
\begin{equation}
R = \sum_{i=2}^n \sum_{j=n-i+2}^n  Y_{ij}.
\label{eq:TotalReserve}
\end{equation}

\begin{table}[!ht]
\begin{center}
\begin{tabular}{c|c c c c c c c}
\toprule
$\frac{\textrm{development year}}{\textrm{accident year}}$&1&2&\ldots&$j$&\ldots&$J-1$&$J$\\
\hline
1&$Y_{11}$&$Y_{12}$&$\ldots$&$Y_{1j}$&$\ldots$&$Y_{1,J-1}$&$Y_{1J}$\\
2&$Y_{21}$&$Y_{22}$&$\ldots$&$Y_{2j}$&$\ldots$&$Y_{2,J-1}$ &\\
$\vdots$&$\ldots$&$\ldots$&$\ldots$&$\ldots$&$\ldots$& &\\
$i$&$Y_{i1}$&$Y_{i2}$&$\ldots$& $Y_{ij}$& & &\\
$\vdots$& $\ldots$&$\ldots$&& & & &\\
$I$&$Y_{I1}$& & & & & & \\
\bottomrule
\end{tabular}
\caption{\label{runoff} Run-off triangle (incremental claim amounts).}
\end{center}
\end{table}

The chain-ladder method uses cumulative data and assumes the existence of development factors $f_j$ for $
j = 2, \ldots, n$ that are estimated by
\[ \hat{f}_j = \frac{\sum_{i=1}^{n-j+1}{C_{ij }}}{\sum_{i=1}^{n-j+1}{C_{i,{j-1}}}}.\]

This leads to the following estimates for the future claim amounts: 
\begin{align*} \hat{C}_{i, n-i+2} & = C_{i,n-i+1} \cdot \hat{f}_{n-i+2} & 1 < i \leqslant n \\ 
\hat{C}_{ik} & = \hat{C}_{i,k-1} \cdot \hat{f}_{k} & 3 \leqslant i \leqslant n \text{  and } n-i+3 \leqslant k \leqslant n. 
\end{align*}

The chain-ladder method is typically specified in the GLM framework using the following multiplicative model for the incremental claim amounts:
\begin{equation}
Y_{ij} \stackrel{i.i.d.}{\sim} \mbox{Poisson}(\zeta_i \xi_j), 
\label{multiplicativestructure}
\end{equation}
with $\zeta_i$ the exposure (i.e. expected claims up to the latest observed development year) of accident year $i$ and $\xi_j$ the expected claims pattern over the different development periods $j$. Since $\zeta_i$ and $\xi_j$ can only be determined up to a constant factor, the condition $\sum_{j=1}^J\xi_J=1$ is often imposed. Formulation (\ref{multiplicativestructure}) has a multiplicative structure for the mean, i.e. $\mu_{ij}=E[Y_{ij}]=\zeta_i\xi_j$ and hence it is straightforward to use a log link function such that
\begin{equation}
\log\left(\mu_{ij}\right)=\tau+\alpha_i+\beta_j=x_{ij}^t\thetab. 
\label{eq:linearstructure}
\end{equation}  
This structure now has a parameter for each row $i$ and each column $j$ and typically the constraint $\alpha_1=\beta_1=0$ is used (to obtain a more convenient interpretation of the parameters). 
Formulations (\ref{multiplicativestructure}) and (\ref{eq:linearstructure}) are equivalent and are reparameterisations of the same structure. The parameters of the first have physical interpretations, whereas the statistical analysis of the latter is more straightforward. The relationship between the parameters and the proof of the equivalence between both models is given in \cite{kremer1982ibnr} and \cite{verrall1991chain}.
The full parameter vector is thus $\thetab = \left(\tau, \alpha_2, \ldots, \alpha_I, \beta_2, \ldots, \beta_I \right)$ and the number of parameters to be estimated equals $2n-1$. The predictor variable $x_{ij}$ consists of the constant $1$ (to include an intercept) and indicators corresponding to each of the rows and columns of the claims triangle except for the first.
By inserting parameter estimates $\hat{\thetab}$ into equation (\ref{eq:linearstructure}) and exponentiating, estimates for the future claim amounts and the total reserve are then obtained.
When estimating the parameters $\thetab$ by maximum likelihood, the results are identical to the estimates obtained by the deterministic chain-ladder method using development factors as shown in \cite{kremer1982ibnr}. For more details we refer to \cite{England:StochClaimsInsurance}, 
\cite{Hoedemakers:GLM} and \cite{Wuthrich:StochasticClaims}.

\subsection{Classical bootstrap procedure}
\label{sec:Bootstrapping}
 
Point estimates 
are very useful of course, but the natural next step is to also estimate their precision. We consider the bootstrap methodology to derive inference corresponding to the point estimates.
To create bootstrap samples we follow the procedure described in \cite{Kaas:ModernAct}, which is based on \cite{England:Analytic,England:StochClaimsInsurance}. 

First, the Pearson residuals, defined as 
\[r_ {ij} = \frac{y_{ij}-\mu_{ij}}{\sqrt{\mu_{ij}}} \qquad \text{for } 1\leqslant i,j\leqslant n \text{ and } i+j\leqslant n+1,\] are calculated by 
replacing the $\mu_{ij}$ by their estimates $\hat{\mu}_{ij}=e^{x_{ij}^t\hat{\thetab}}$.
When the number of observations is not very large, the residuals suffer from a small sample bias. To remedy this, \cite{England:Analytic} and \cite{England:AnalyticAdd} propose to adjust the residuals $\boldsymbol{r}=\{r_{ij}|1\leqslant i,j\leqslant n; i+j\leqslant n+1\}$ by multiplying them by a correction factor:
\[\boldsymbol{r}^E = \sqrt{\frac{N}{N-p}} \boldsymbol{r},\]
where $N=n(n+1)/2$ is the sample size and $p = 2n-1$ is the number of fitted parameters. 
\cite{Pinheiro:BootstrapClaims} argue that it is better to use an individual adjustment factor for each residual $\{r_k|k=1,\ldots, N\}=\{r_{ij}|1\leqslant i,j\leqslant n; i+j\leqslant n+1\}$ as opposed to a global correction factor. In correspondence with the classical linear regression model they propose to use the hat matrix $H$ of the model to standardize the Pearson residuals as follows
\[ r_{ij}^P = \frac{r_{ij}}{\sqrt{1-h_{kk}}}\] where $h_{kk}$ is the corresponding element on the diagonal of the hat matrix.
For a Poisson GLM, the hat matrix is given by $H = W^{1/2}X(X^TWX)^{-1}X^TW^{1/2}$, where $W$ is a diagonal matrix with elements $\{\mu_{ij}|1\leqslant i,j\leqslant n; i+j\leqslant n+1\}$ on the diagonal (see e.g. \cite{McCullagh:GLM} for more details). 
\cite{Cordeiro_PResidGLM} recently provided general expressions for the expectation and variance of Pearson residuals in the GLM framework up to the first order $O(N^{-1})$. Using equations (\ref{eq:PoissonGLM_Var}) and (\ref{eq:PoissonGLM_Exp}), calculated in Appendix \ref{sec:Cordeiro},  we obtain that 
\begin{eqnarray*}
 E[\boldsymbol{r}] &=& \mbox{diag}\left(-\frac{1}{2}(I-H)HW^{-1/2}\right)\\
\var[\boldsymbol{r}] &=& \mbox{diag}\left(I - H\right)
\end{eqnarray*}
where `$\mbox{diag}$' indicates that we extract the diagonal elements of a matrix. These correction terms lead to the following adjusted residuals \[\boldsymbol{r}^C = \frac{\boldsymbol{r} - E[\boldsymbol{r}]}{\sqrt{\var[\boldsymbol{r}]}}.\] 
Replacing the parameters $\mu_{ij}$ in the expressions of $H$ and $W$ by their estimates $\hat{\mu}_{ij}$ yields the expectation and variance correction term of the sample Pearson residuals.

The bootstrap methodology can be applied by using the original Pearson residuals or any of the above explained adjustments. 
From now on, we use the general notation $\boldsymbol{r}^*$ for the residuals of choice, i.e. $\boldsymbol{r}^*$ is any of $\boldsymbol{r},\boldsymbol{r}^E,\boldsymbol{r}^P$ or $\boldsymbol{r}^C$. In the next sections, all results shown are obtained for $\boldsymbol{r}^*=\boldsymbol{r}^C$, but the conclusions also hold for the other adjustment options.

After obtaining the residuals, the following bootstrap steps are performed many times, e.g. for $b=1, \, \ldots,$ 10,000:
\begin{enumerate}
	\item Resample with replacement from the residuals $\boldsymbol{r}^*$ leading to a new vector of residuals $\boldsymbol{r}^{(b)}$. Since there is only one observation for the last accident year and the last development year, the corresponding residuals (on the cornerpoints of the run-off triangle) will be exactly zero when fitting a GLM. As suggested by \cite{England:AnalyticAdd} and \cite{Pinheiro:BootstrapClaims}, these two residuals are not resampled (whereas all other residuals have equal resampling probabilities).
	\item Backtransform the new residuals $\boldsymbol{r}^{(b)}$ in order to obtain a new pseudo-history: 
				\[ \boldsymbol{y}^{(b)} = \boldsymbol{r}^{(b)} \sqrt{\boldsymbol{\mu}} + \boldsymbol{\mu}.\]
	\item Refit the GLM on the new pseudo-history $\boldsymbol{y}^{(b)}$ in order to obtain a new total future reserve estimate $R^{(b)}$. 
\end{enumerate}

\section{Robust chain-ladder method and classical bootstrapping}
\label{sec:RobchainladderBootstrapping}

The robust chain-ladder method of \cite{Verdonck:IFClaims} is briefly described and the effect of outliers on the total reserve estimates obtained by the classical and robust chain-ladder method is illustrated on the well studied claims data set of \cite{Taylor:Data} as well as on a simulated data set.  For both estimators we also evaluate the effect of outliers on an estimated upper limit for the total reserve obtained by using the classical bootstrap procedure. 

\subsection{Robust chain-ladder method}
\label{sec:RobustGLMEstimator}

\cite{Verdonck:IFClaims} already showed the sensitivity of the classical 
chain-ladder method to outliers and therefore proposed a robust alternative, which is based 
on a general class of $M$-estimators of Mallow's type proposed by \cite{Cantoni:genreg}, with the following set of estimating equations: 
\begin{equation}
\label{eq:CantoniPoisGLM}
\sum_{i=1}^n\sum_{j=1}^{n-i+1}{\boldsymbol{\psi}(y_{ij}, \mu_{ij})} = \sum_{i=1}^n\sum_{j=1}^{n-i+1}\left[ \psi_c(r_{ij})w(x_{ij})\frac{1}{V^{1/2}(\mu_{ij})}\mu_{ij}'-a(\boldsymbol{\thetab}) \right]=0.
\end{equation}
The vector $\thetab = \left(c, \alpha_2, \ldots,\alpha_I, \beta_2, \ldots, \beta_I \right)$ as in equation (\ref{eq:linearstructure}) and $\mu_{ij}'$ denotes the derivative of $\mu_{ij}$ with respect to $\thetab$.
The constant 
$a(\boldsymbol{\thetab})=\frac{2}{n(n+1)}
\sum_{i=1}^n\sum_{j=1}^{n-i+1} E \left[\psi_c(r_{ij})\right]\omega(x_{ij})
\frac{1}{V^{1/2}(\mu_{ij})}\mu_{ij}'$
ensures Fisher consistency and
 \begin{equation*}
\psi_c(r_{ij})=\left\{ \begin{array}{ll}
r_{ij} & |r_{ij}| \leqslant c\\
c \cdot \textrm{sign}(r_{ij})& |r_{ij}|\geqslant c,
\end{array}
\right.
\end{equation*}
is the Huber function. The weight function $w(x_{ij})$ gives less weight to observations with outlying predictors, whereas the Huber function limits the effect of observations with an outlying response by truncating the residual. Therefore, the influence function of deviations on the response and on the predictors is bounded separately.
Note that for a log-link 
Poisson GLM one obtains $V^{1/2}(\mu_{ij}) = \sqrt{\mu_{ij}}$.
\cite{Cantoni:genreg} showed that the estimating equations may 
equivalently be written as 
\begin{equation}
\label{eq:Cantoni_Eq}
\sum_{i=1}^n\sum_{j=1}^{n-i+1}\left[ \left(\psi_c(r_{ij}) - E \left[\psi_c(r_{ij})\right] \right) w(x_{ij}) \sqrt{\mu_{ij}} x_{ij} \right]=0
\end{equation}
with 
\begin{align}
\label{eq:Cantoni_ExpHuberRes}
E \left[\psi_c(r_{ij})\right] = & c \left[ \, \Prob{Y_{ij} \geqslant j_2 + 1} -
 \Prob{Y_{ij} \leqslant j_1}  \right]  \\ &+ \sqrt{\mu_{ij}} 
\left[ \Prob{Y_{ij} = j_1} - \Prob{Y_{ij} = j_2}\,\right]\nonumber
\end{align}
and $j_1 = \floor{\mu_{ij} - c\sqrt{\mu_{ij}}}$ and 
$j_2 = \floor{\mu_{ij} + c\sqrt{\mu_{ij}}}$.

\subsection{Illustration of the effect of outlier(s)}
\label{sec:IllustrationOfMethodologyAndTheNeedForFurtherSteps}
\begin{figure}[!h]
			\centering
			\includegraphics[width=0.75\textwidth]{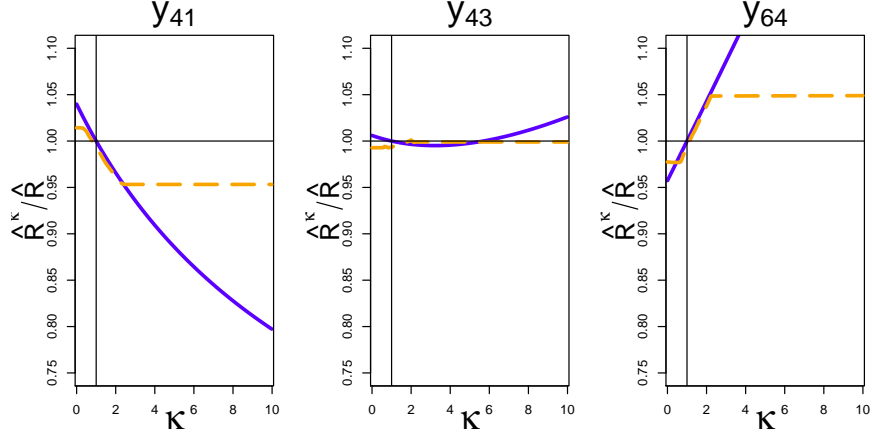}
						\caption{Influence of $\kappa$ on the ratio between the original total reserve estimate, using $y_{ij}$, and the total reserve estimate corresponding to $\kappa$, using $\kappa y_{ij}$ for the classical (blue solid line) and the robust (orange dashed line) chain-ladder method. Left, middle and right plot correspond respectively to $y_{41}, y_{43}$ and $y_{64}$.}
			\label{fig:sensitivity}
\end{figure}

In this section we illustrate the classical and robust techniques on the widely studied data set of \cite{Taylor:Data}, shown in table \ref{engver}.
\addtolength{\tabcolsep}{-1mm}
\begin{table}[!ht]
\footnotesize
\begin{center}
\begin{tabular}{c|r r r r r r r r r r}
\toprule
			&1&2&3&4&5&6&7&8&9&10\\
			\hline
		 	1 &357,848	&766,940		&610,542		&482,940		&527,326		&574,398 	&146,342   &139,950  &227,229  &67,948\\
			2 &352,118	&884,021		&933,894		&1,183,289	&445,745		&320,996		&527,804   &266,172  &425,046  & \\
			3 &290,507	&1,001,799	&926,219		&1,016,654  &750,816   &146,923   &495,992   &280,405  &        & \\
			4 &310,608	&1,108,250	&776,189		&1,562,400	&272,482		&352,053   &206,286   &        &        & \\
			5 &443,160	&693,190		&991,983		&769,488   &504,851   &470,639   &         &        &        & \\
			6 &396,132	&937,085		&847,498		&805,037	  &705,960		&         &         &        &        & \\
			7 &440,832	&847,631		&1,131,398	&1,063,269  &         &         &         &        &        & \\
			8 &359,480	&1,061,648	&1,443,370	&   			&					&         &         &        &        & \\
			9 &376,686	&986,608		&					&         &         &         &         &        &        & \\
			10&344,014	&					&					&         &         &         &         &        &        & \\
			\bottomrule
\end{tabular}
\end{center}
\caption{\label{engver} claims data from Taylor and Ashe (1983).}
\end{table}
\addtolength{\tabcolsep}{1mm}
The classical and robust total reserve estimates equal respectively $18,680,856$ and $18,562,327$. To study the effect of an outlier on the classical and robust chain-ladder method, one observation $y_{ij}$ is adjusted to $\kappa y_{ij}$ for $\kappa \in [0,10]$. We calculate the total reserve estimates before and after this adjustment, respectively denoted by $\hat{R}$ and $\hat{R}^{\kappa}$.
For observations $y_{41}$, $y_{43}$ and $y_{64}$, we plot the ratio $\frac{\hat{R}^{\kappa}}{\hat{R}}$ in Figure \ref{fig:sensitivity}.
It is immediately clear that the influence of an outlier on the robust chain-ladder method (orange dashed line) is very limited in all cases, even if $\kappa$ becomes very large. On the other hand, the outlier has a large, unbounded effect on the classical chain-ladder method (blue solid line). Depending on the location of the outlier in the run-off triangle, the estimated reserve $\hat{R}^{\kappa}$ can largely increase or decrease. For example, an unusually large claim amount in the past ($\kappa>1$) cannot only result in a much bigger reserve estimate, but may also result in a reserve estimate that is much smaller even when no anomalies occur in the future anymore. This illustrates the sensitivity of the classical methodology and hence, there is no guarantee that the estimated classical reserve suffices to cover future claim amounts.

\begin{figure}[!ht]
      \centering
      \includegraphics[width=0.6\textwidth]{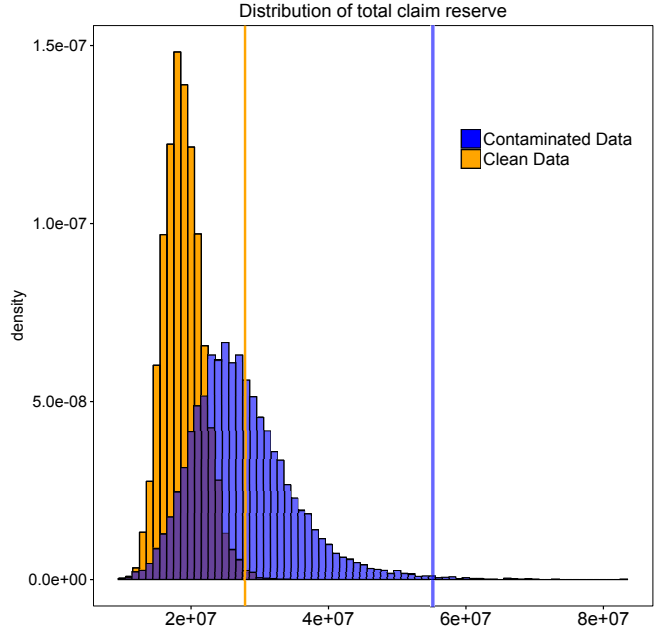}
			\caption{Distribution of bootstrap total reserve estimates based on the classical chain-ladder for the data of \cite{Taylor:Data}. total reserve estimates based on the clean data are shown in orange (light color) while total reserve estimates based on the contaminated data are shown in blue (dark color). Vertical lines indicate the corresponding 99.5$\%$ quantiles.  }
			\label{fig:TaylorAshe_Illustrate_Conta}
\end{figure}

We now investigate the effect of an outlier on upper limits for the total reserve estimated by the classical bootstrap procedure. For  the 
classical chain-ladder the distribution of the obtained classical bootstrap total reserve estimates is shown in orange (light color)
in Figure \ref{fig:TaylorAshe_Illustrate_Conta}. The 99.5$ \%$ 
quantile of this bootstrap distribution corresponds to a total reserve estimate of approximately $27.8$ million (indicated by a vertical line). To illustrate the effect of 
an outlier, we multiply observation $y_{27}$ in Table~\ref{engver} by ten. From the 
corresponding bootstrap distribution, which is shown in blue (dark color) in Figure 
\ref{fig:TaylorAshe_Illustrate_Conta}, it is seen that now a $99.5\% $ quantile of roughly 
55 million is obtained. This clearly indicates the sensitivity of the 
classical chain-ladder method in combination with the classical bootstrap 
procedure. 

\begin{figure}[!ht]
			\centering
			\includegraphics[width=0.6\textwidth]{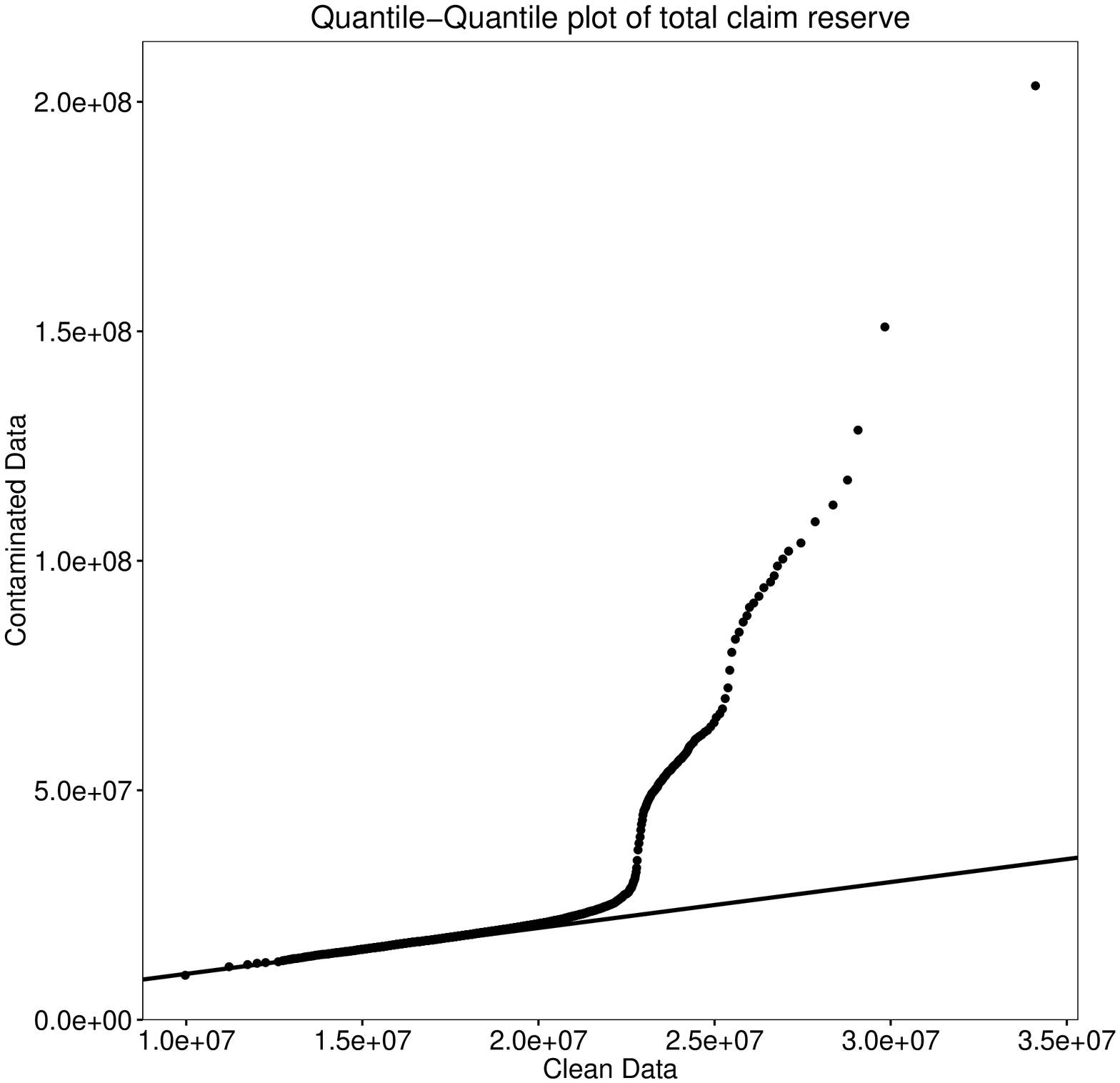}
						\caption{Quantile-quantile plot of total claim reserve distribution comparing the traditional procedure on clean data with the robust chain-ladder method combined with the classical bootstrap technique on contaminated data.}
			\label{fig:TaylorAshe_Ill_RobConta_Fails}
\end{figure}

To limit the effect of outlying observations in the obtained bootstrap distribution, we now combine the robust chain-ladder method with the traditional bootstrap procedure and apply it to the contaminated data set of \cite{Taylor:Data} (i.e. $y_{27}$ is multiplied by ten). Note that for the adjusted residuals $\boldsymbol{r}^P$ and $\boldsymbol{r}^C$, we need the hat matrix corresponding to the robust Poisson GLM estimator of \citet{Cantoni:genreg}. This expression is derived in Appendix \ref{sec:hatRobGLM}. To compare the distribution of the obtained total reserves for the contaminated data with the previously obtained \textit{original} reserve estimates (i.e. the reserves obtained by bootstrapping the 
classical chain-ladder method on the non-contaminated data), the quantile-quantile plot is given in Figure \ref{fig:TaylorAshe_Ill_RobConta_Fails}. As expected, a large range of the obtained quantiles are now
relatively close to the corresponding original quantiles (since we plugged in the 
robust chain-ladder method). However, the upper quantiles of the robust chain-ladder method on contaminated data deviate significantly from their original counterparts. This is caused by the resampling with replacement in the bootstrap process. During the resampling process, it 
might occur that the outliers are selected so many times in a bootstrap sample that 
the robust method can no longer withstand this huge amount of contamination. 
This will happen several times if we generate 10,000 resamples, yielding unreliable bootstrap estimates that affect the bootstrap distribution, especially if one is interested in an upper quantile that lies far in the tail of the distribution. 

This sensitivity of the bootstrap technique, even in combination with robust 
point estimators, is well-known in the robust statistics literature and was 
first discussed by \cite{ Stromberg:BootstrapBreakdown}. This seminal paper 
gave rise to the development of several robust alternatives for the classical bootstrap. In the next section we discuss the robust bootstrap methods that have been proposed (up to our knowledge) and we adapt them to the claims reserving framework.

To end this section, we compare the outlier detection capacity of the classical and robust chain ladder on the simulated data triangle in Table \ref{table:IllRes} which was generated from the benchmark model (see section \ref{sec:BenchmarkModel}). In this triangle we replaced observations $y_{16}, y_{36}, y_{61}, y_{65}$ and $y_{24}$ respectively by 33,000, 35,000, 120,000 and 65,000 and 7000. The classical and robust total reserve estimates are $314,240$ and $155,086$, respectively. Hence, inspection of the data is needed to understand this difference. In practice, one often tries to detect outliers using diagnostics starting from a fit obtained by the classical estimation method. The Pearson residuals obtained from the classical chain-ladder method are given in Table \ref{table:IllRes_Pearson}. Besides both corner points and $r_{51}$, all residuals have an absolute value much larger than 3 (which is typically used as a benchmark in a classical outlier detection rule). Apparently, the classical method is so heavily affected by the outliers that the fitted model does not allow to detect the deviating observations anymore. For example, all observations in row two and six and column six have become inflated, making it impossible to flag observation $y_{24}$ as an outlier. This is called the \textit{masking} effect. On the other hand, observation $y_{62}$ which is a regular observation yields the largest residual and hence appears to be the most outying observation. This effect is known as \textit{swamping}. To avoid these adverse effects, the goal of robust statistics is to obtain a fit which is similar to the fit we would have found without outliers in the data. An important benefit is that it then becomes possible again to detect the outliers by their large deviation from the robust fit. The robust chain-ladder method automatically provides a robustness weight for each observation, which is given by 
\begin{equation}\label{eq:robweights} \frac{\psi_c(r_{ij})}{r_{ij}}. \end{equation}
If a residual $r_{ij}$ is large compared to its Huberized value $\psi_c(r_{ij})$, then the observation receives a low weight and is identified as a potential outlier. The robustness weights obtained by fitting the robust chain-ladder method to the data are shown in Table \ref{table:IllResWeights}. The outlying cells are now correctly flagged by the robustness weights and can be studied in more detail. 

\begin{table}[ht]
\scriptsize
\begin{center}
\begin{tabular}{c|r r r r r r r r r r}    
\toprule
 & 1 & 2 & 3 & 4 & 5 & 6 & 7 & 8 & 9 & 10 \\ \hline
  1 & 9891 & 7914 & 6092 & 4842 & 3812 & 3012 & 2370 & 1842 & 1522 & 1139 \\ 
  2 & 10373 & 8210 & 6624 & 5304 & 4146 & 3218 & 2486 & 1939 & 1525 &  \\ 
  3 & 10890 & 8611 & 6702 & 5345 & 4119 & 3328 & 2502 & 2054 &  &  \\ 
  4 & 11463 & 9036 & 7091 & 5596 & 4316 & 3429 & 2783 &  &  &  \\ 
  5 & 12050 & 9385 & 7424 & 5828 & 4425 & 3533 &  &  &  &  \\ 
  6 & 12675 & 9793 & 7594 & 6026 & 4623 &  &  &  &  &  \\ 
  7 & 13116 & 10185 & 7928 & 6315 &  &  &  &  &  &  \\ 
  8 & 13473 & 10652 & 8234 &  &  &  &  &  &  &  \\ 
  9 & 14205 & 11101 &  &  &  &  &  &  &  &  \\ 
  10 & 14598 &  &  &  &  &  &  &  &  &  \\ 
	\bottomrule
\end{tabular}
\end{center}
\caption{\label{table:IllRes} Simulated run-off triangle.}
\end{table}

\begin{table}[ht]
\scriptsize
\begin{center}
\begin{tabular}{c|r r r r r r r r r r}    
\toprule
 & 1 & 2 & 3 & 4 & 5 & 6 & 7 & 8 & 9 & 10 \\ \hline
  1 & -62.0 & 8.4 & 10.3 & 9.9 & -61.1 & 97.9 & -11.6 & -6.5 & -7.8 & -0.0 \\ 
  2 & -12.0 & 53.3 & 55.0 & 83.4 & -27.3 & -82.3 & 12.9 & 15.4 & 9.8 &  \\ 
  3 & -63.3 & 9.2 & 12.0 & 11.8 & -63.7 & 96.7 & -13.3 & -5.6 &  &  \\ 
  4 & -3.7 & 63.8 & 61.5 & 54.9 & -26.3 & -81.6 & 18.8 &  &  &  \\ 
  5 & -0.6 & 66.7 & 65.2 & 57.6 & -26.4 & -82.2 &  &  &  &  \\ 
  6 & 120.9 & -129.2 & -109.2 & -96.5 & 89.1 &  &  &  &  &  \\ 
  7 & -46.3 & 28.1 & 29.1 & 27.0 &  &  &  &  &  &  \\ 
  8 & -41.3 & 35.6 & 35.0 &  &  &  &  &  &  &  \\ 
  9 & -29.3 & 46.7 &  &  &  &  &  &  &  &  \\ 
  10 & 0.0 &  &  &  &  &  &  &  &  &  \\ 
	\bottomrule
\end{tabular}
\end{center}
\caption{\label{table:IllRes_Pearson} Pearson residuals obtained after fitting a Poisson GLM to the data in Table \ref{table:IllRes}.}
\end{table}

\begin{table}[ht]
\scriptsize
\begin{center}
\begin{tabular}{c|r r r r r r r r r r}    
\toprule
 & 1 & 2 & 3 & 4 & 5 & 6 & 7 & 8 & 9 & 10 \\ \hline
  1 & 1.00 & 1.00 & 1.00 & 1.00 & 1.00 & \cellcolor{red!100}0.00 & 1.00 & 1.00 & 1.00 & 1.00 \\ 
  2 & \cellcolor{red!30}0.70 & 1.00 & \cellcolor{red!19}0.81 & \cellcolor{red!95}0.05 & \cellcolor{red!27}0.73 & 1.00 & 1.00 & 1.00 & 1.00 &  \\ 
  3 & 1.00 & 1.00 & 1.00 & 1.00 & 1.00 & \cellcolor{red!100}0.00 & \cellcolor{red!10}0.90 & 1.00 &  &  \\ 
  4 & 1.00 & 1.00 & 1.00 & 1.00 & 1.00 & 1.00 & 1.00 &  &  &  \\ 
  5 & 1.00 & 1.00 & 1.00 & 1.00 & \cellcolor{red!25}0.75 & \cellcolor{red!24}0.76 &  &  &  &  \\ 
  6 & \cellcolor{red!100}0.00 & 1.00 & 1.00 & 1.00 & \cellcolor{red!100}0.00 &  &  &  &  &  \\ 
  7 & 1.00 & 1.00 & 1.00 & 1.00 &  &  &  &  &  &  \\ 
  8 & 1.00 & 1.00 & 1.00 &  &  &  &  &  &  &  \\ 
  9 & 1.00 & 1.00 &  &  &  &  &  &  &  &  \\ 
  10 & 1.00 &  &  &  &  &  &  &  &  &  \\ 
	\bottomrule
\end{tabular}
\end{center}
\caption{\label{table:IllResWeights} Robust weights obtained after fitting a robust Poisson GLM to the data in Table \ref{table:IllRes}.}
\end{table}

\section{Robust bootstrap methods}
\label{sec:RobustBootstrapMethods}

To obtain robust inference, \cite{Stromberg:BootstrapBreakdown} proposed to use robust measures of 
location and scale of the classical bootstrap distribution corresponding to the robust estimators. Typically, the median and the median absolute 
deviation (see \cite{Hampel:IC}) are applied instead of the classical mean 
and standard deviation. Obviously,  we cannot use this method to calculate quantiles of the bootstrap distribution. \cite{ Stromberg:BootstrapBreakdown} also proposed the 
deleted-$d$ jackknife procedure, in which a bootstrap sample consists of $N-d$ 
data points instead of $N$ data points to increase the probability of sampling a clean 
subset. However, in our framework this poses problems as the resampled triangle would then be incomplete. 

On the other hand, \cite{ Singh:BootstrapBreakdown} proposed 
to robustify the bootstrap procedure itself by resampling from a winsorized 
version of the original sample. In a regression context, one may resample either the residuals or the original observations themselves. Due to the specific structure of a claims triangle, we resample the residuals. To winsorize the residuals the quantiles $Q_c$ and $Q_{1-c}$ of the residuals are determined for $c 
\in \left]0,0.5\right[ $. The $c$ percent smallest residuals are then replaced 
by $Q _c$ whereas the $c$ percent largest ones are replaced by $Q_{1-c}$. 

More recently, \cite{Salibian:BootstrapRobReg} and \cite{Salibian:FRB} 
proposed a fast and robust bootstrap method that is asymptotically consistent under reasonable 
regularity conditions.  The corresponding R-package (\cite{VanAelst:RFRB}) considers three multivariate settings: 
principal component analysis, Hotelling tests and multivariate regression. 
Another recent alternative is proposed by \cite{Amado:IFB} and \cite{Amado:IFB2}.
They suggest to resample observations with different probabilities such 
that potentially harmful observations have low sampling probability. They propose to determine these probabilities based on the influence function of the estimator.  We now 
discuss both state-of-the-art robust bootstrap methods in more detail and 
describe their implementation in the claims reserving framework.

\subsection{Fast and Robust Bootstrap} 
\label{sec:FastAndRobustBootstrap} 

For each pseudo-history, most robust bootstrap procedures need to recalculate the robust estimator, which in itself is 
already computationally intensive compared to the classical estimator. 
An important aim of the fast and robust bootstrap (\mbox{FRB}) is 
to reduce this computational load \citep{Salibian:BootstrapRobReg,Salibian:FRB}.

Within the context of claims reserving the \FRB procedure can be described as 
follows.  Let $\thetab \in \mathbb{R}^{2n-1}$ be the parameter of interest.
Note that  $\thetab$ consists of the intercept $c$ and the parameters $\alpha_i$ and $\beta_j$ for
$2 \leqslant i,j \leqslant n$, as before. 
Now, suppose that $\hat{\thetab}$
 is a solution of a fixed-point equation $\hat{\thetab} =  g_N(\hat{\thetab}
)   $ where $g_N$ depends on the sample of size $N=n(n+1)/2$ composed of the 
historical claims data. Similarly, for a resampled pseudo-history the corresponding estimator $\hat{\thetab}^{(b)}$  
solves the equation $\hat{\thetab}^{(b)} = g^{(b)} _N(\hat{\thetab}^{(b)} )$ where 
the function  $g^{(b)}_N$ now depends on the bootstrap sample. As a first step 
one may consider $\tilde{\thetab}^{(b)} = g ^{(b)}_N(\hat{\thetab}) $  
as an approximation for the bootstrap replicate $\hat{\thetab}^{(b)}$ of the estimator $ \hat{\thetab}$. This simple procedure avoids having to recompute the robust estimator for each bootstrap sample, which may be 
computationally hard. However, these approximate bootstrap replicates typically 
underestimate the true variability since all bootstrap approximations $\tilde{\thetab}^{(b)}$ are based on 
the same initial estimate $\hat{\thetab}$. Therefore, \cite{Salibian:FRB} 
proposed a linear correction resulting in the FRB bootstrap replicates 
\begin{equation*}
\hat{
\thetab}^{(b)}_{\text{FRB}} = \hat{ \thetab} + \left[I_p - \nabla g_N(\hat{\thetab}) \right]^{
-1} \cdot \left( g ^{(b)}_N(\hat{\thetab}) -  \hat{\thetab} \right),
\label{eq:FRB}
\end{equation*}  where 
$\nabla g_N(\hat{\thetab})$ corresponds to the gradient of $g_N$ evaluated at $\hat{\thetab}$ and $I_p$ denotes the identity matrix of dimension $p = 2n-1$.

This procedure was later extended by \cite{Camponovo:RobustSubsampling} 
to accommodate $M$-estimators defined by the solution of equations of the 
following type \[ \psi_N(\hat{\thetab}) = \sum_{k=1 }^{N}{\psi(\hat{\thetab})} = 0.\] 
Fast and robust bootstrap estimates are then obtained as follows: 
\begin{equation*}
\hat{\thetab}^{(b)}_{\text{FRB}} = \hat{ \thetab} - \left[\nabla \psi_N(\hat{\thetab})\right]^{
-1} \cdot \psi^{(b)}_N(\hat{\thetab}).
\label{eq:FRB_GLMPois}
\end{equation*}

The \FRB paradigm can be used for the Poisson GLM estimator of \cite{Cantoni:genreg}. Based on the estimating equations (\ref{eq:Cantoni_Eq}) we define
\begin{equation}
\label{eq:FRB_Def_xi}
\psi_N(\hat{\thetab}) = \sum_{i=1}^n \sum_{j=1}^{n-i+1} \left[ \left(\psi_c(r_{ij}) - E \left[\psi_c(r_{ij})\right] \right) w(x_{ij}) \sqrt{\mu_{ij}} x_{ij} \right].
\end{equation}

To calculate $\nabla \psi_N(\hat{\thetab})$ we readily obtain
\[\frac{\partial}{\partial\thetab}{\left( \sqrt{\mu_{ij}} \right)} = \frac12 \mu_{ij}^{1/2}x_{ij} \quad \text{ and } \quad \frac{\partial}{\partial\thetab}{\left(\psi_c(r_{ij})\right)} = -\frac12 I_{|r_{ij}|\leqslant c} \left(\frac{y_{ij}+\mu_{ij}}{\mu_{ij}^{1/2}} \right) x_{ij}, \]
with $I$ the indicator function. It then follows that 
\begin{align*}
\nabla \psi_N(\hat{\thetab}) = \sum_{i=1}^n\sum_{j=1}^{n-i+1} w(x_{ij}) x_{ij} & \left[ \left( -\frac12 I_{|r_{ij}|\leqslant c} \left(\frac{y_{ij}+\mu_{ij}}{\mu_{ij}^{1/2}} \right) x_{ij} - \frac{\partial}{\partial\thetab} \left( E \left[\psi_c(r_{ij})\right] \right) \right) \sqrt{\mu_{ij}} \right. \\
& +  \left. \left(\psi_c(r_{ij}) - E \left[\psi_c(r_{ij})\right]\right) \frac12 \mu_{ij}^{1/2}x_{ij} \right]^t. 
\end{align*}
The term $E \left[\psi_c(r_{ij})\right]$ however is not differentiable when using its 
exact expression. Therefore, we use the $\chi^2$ distribution as an 
approximation for the Poisson distribution and afterwards apply the Wilson-Hilferty 
approximation. The detailed derivation can be found in Appendix \ref{sec:FRB}.

\subsection{Influence Function Bootstrap}
\label{sec:InfluenceFunctionBootstrap}
 \cite{Amado:IFB} 
propose to determine resampling probabilities for the observations by using the influence function corresponding to the classical estimator. The definition of the influence 
function is based on the concept of a statistical functional, which we recall first.
 Let $T_N$ denote a $p$-dimensional estimator defined for any sample 
size $N$, with $N=(n+1)n/2$ and $p=2n-1$ in our setting. A statistical functional $T$ corresponding to the estimator $T_N$ is a map defined on  $p$-variate distributions $G$ on $\mathbb{R}^p$ such 
that $ T(G_N(\{y_{ij}|1\leqslant i,j\leq n; i+j\leqslant n+1 \}) = T_N(\{y_{ij}|1\leqslant i,j\leq n; i+j\leqslant n+1 \})$ where $G_N$ is the empirical distribution of the sample $\{y_{ij}|1\leqslant i,j\leq n; i+j\leqslant n+1 \}$. 

Let $\{Y_{ij}\,|\,1\leqslant i,j\leqslant n; i+j\leq n+1\}\sim F$ and assume that the marginal distribution of $Y_{ij}$, denoted by $F_{\mu_{ij}}=F_{i,j}$, is Poisson distributed with mean $\mu_{ij}$, see equation \eqref{eq:linearstructure}. For any $1\leqslant l,m \leqslant n$ with $l+m \leqslant n+1$ and $z \geqslant 0$, we define $F_{l,m,\varepsilon,z}$ such that 
\begin{displaymath}
\begin{cases}
\label{contdis}
Y_{ij} \sim F_{i,j} \qquad  \qquad \forall (i,j)\neq(l,m)\\
Y_{l,m} \sim (1-\varepsilon)F_{l,m}+\varepsilon \Delta_z 
\end{cases}
\end{displaymath}
where $\Delta_z$ is a Dirac measure putting all its mass at value $z$. 

The influence function of the functional $T$ at a distribution $F$  is then defined as \[\mbox{ IF}([z,l,m], T, F) = \lim_{\varepsilon 
\downarrow 0}{\frac{T(F_{l,m,\varepsilon,z}) - T (F) }{\varepsilon}}.\] 
Intuitively, the influence function measures the change of the estimator when 
the model distribution is perturbed by an infinitesimal small amount of 
contamination at location $z$. 

\cite{Amado:IFB} propose to measure the effect of each data point on the parameter
estimates by estimating the value of the population influence function at each observation using estimates of the model parameters based on the sample. However, if a non-robust 
estimator is employed then an outlier may already have such a high effect on the sample estimates themselves that it results in a low value for the influence 
function evaluated at the outlier. This phenomenon is known as masking in the
robust statistics literature. Conversely, the influence function of a robust 
estimator will also have small values at the outliers, since 
outliers should have a low effect on robust estimators \citep{Pison:DiagPlot}. Therefore, \cite{ 
Amado:IFB} consider the influence function of the classical non-robust 
estimator (denoted by $T^{nr}_{\thetab}$) using parameter estimates $\hat{\thetab}^r$ obtained by a robust estimator, i.e. the influence function is evaluated at $F_{\hat{\thetab}^r}$.
By combining robust parameter estimates with the influence 
function of the non-robust estimator, its values for outliers should be very 
high compared to regular observations. 
This idea is formalized by the Robust Empirical Standardized Influence 
Function (RESIF) that is defined as: 
\begin{equation}
\label{eq:RESIFDef}
 \mbox{RESIF}([z,l,m],T_{\thetab}^{nr},F_{\hat{\thetab}^{r}}) = \left[ \mbox{IF}([z,l,m],T_{\thetab}^{nr},F_{\hat{\thetab}^r})^{t} \cdot \mbox{ V}^{-1}(T_{\thetab}^{nr},F_{\thetab^r}) \cdot \mbox{IF}([z,l,m],T_{\thetab}^{nr} ,F_{\hat{\thetab}^r}) \right]^{\frac{1}{2}}
\end{equation}
where 
\[ \mbox{V}(T_{\thetab}^{nr} ,F_{\hat{\thetab}^r}) = \mbox{E}_{F_{\hat{\thetab}^r}}\left[ \mbox{IF}([z,l,m],T_{\thetab}^{nr} ,F_{\hat{\thetab}^r}) \cdot \mbox{IF}([z,l,m],T_{\thetab}^{nr},F_{\hat{\thetab}^r})^{t} \right]. \] 

Based on the $\mbox{RESIF}$ function the bootstrap procedure is adjusted. We now discuss this procedure in more detail in the context of claims reserving.
\begin{enumerate}
	\item Calculate the values: 
				\[\mbox{RESIF}_{ij} = \mbox{RESIF}([y_{ij},i,j],T_{\thetab}^{nr},F_{\hat{\thetab}^{r}})  \]
				for each of the observations in the sample $\{y_{ij}|1\leqslant i,j\leq n; i+j\leqslant n+1 \}$.
	\item Choose constants $d,\gamma$ and $c$ and calculate for each observation $y_{ij}$ a weight $w_{ij}$  as follows:
				\begin{align*}
				w_{ij} & = I_{[0,c]}\left( \,|\mbox{RESIF}_{ij}| \, \right) \\
				& + I_{]c,\infty[}\left( \,|\mbox{RESIF}_{ij}| \, \right) \cdot \eta_{d,\gamma}\left(c, |\mbox{RESIF}_{ij}| \right)
				\end{align*} 
				where 
				\begin{equation}
				\eta_{d,\gamma}(c,x)=\left[ 1+\frac{(x-c)^2}{\gamma d^2}\right]^{-\frac{\gamma+1}{2}}
								\label{weightIFB}
				\end{equation}
				and $I_{[0,c]}$ indicates the indicator function on the interval $[0,c]$.
	\item Compute the resampling probability
				\[p_{ij} = \frac{w_{ij}}{\sum_{k=1}^n\sum_{\ell=1}^{n-i+1}{w_{k\ell}}} \quad 1\leqslant i,j \leqslant n \text{ and } i + j \leqslant n + 1 \] 
				for each observation.
\end{enumerate}
The bootstrap procedure is then initiated by resampling the residuals $\boldsymbol{r}^*$ using these resampling probabilities. Note that the residuals of the two corner points are still not resampled. 

The RESIF for the Poisson GLM estimator of \cite{Cantoni:genreg} is derived in Appendix \ref{sec:ProofOfTheorem} and the result is given in Theorem \ref{theo:SIF}. 

\begin{theorem}
\label{theo:SIF}
The Robust Empirical Standardized Influence Function of the M-estimator defined by equation \eqref{eq:FRB_Def_xi} for the Poisson GLM is given by 
\[ \mbox{RESIF}([z,l,m],T_{\thetab}^{nr},F_{\hat{\thetab}^{r}}) = \left[\frac{(z-\hat{\mu}_{lm})}{\hat{\mu}_{lm}} \right]^{1/2} \quad 1 \leqslant l,m \leqslant n \text{ and } l+m \leqslant n+1 .\]
\end{theorem}

\cite{Amado:IFB} propose to apply the classical GLM estimator to calculate the parameter estimates corresponding to  the bootstrap samples. This drastically reduces the computation time that is needed for the bootstrap. The robust initial estimate combined with the weights 
based on the influence function protect against outliers in the bootstrap samples. The resulting 
procedure is both fast and robust. Note that the consistency of this methodology has not been formally proven.

\section{Simulation Study}
\label{sec:SimulationStudy}

To compare the different bootstrap procedures, their performance is evaluated in an extensive simulation study. Three different models to generate complete run-off triangles (lower and upper part) are considered. 
We start with a simple benchmark model already allowing us to study several properties of the various bootstrap methods. Then, we turn our attention to more realistic simulation models from the claims reserving literature. 
To simulate run-off triangles in the claims reserving framework we focus on the models proposed by \cite{Schiegl:Handbook} and \cite{Cowell:Bootstrap}.
Before discussing the simulation setup and results, we briefly describe the three data generating models and their corresponding parameters. Values for these parameters were chosen to obtain realistically looking data sets (based on the proposed values in \cite{Schiegl:Handbook} and \cite{Cowell:Bootstrap}).
Note that other constants than below were also studied, but they led to similar results and conclusions and hence were omitted. 

\subsection{Data generating models}
\label{sec:DataModels}

\subsubsection{Benchmark model}
\label{sec:BenchmarkModel}
The benchmark model is a simplification of the model proposed by \cite{Schiegl:Handbook}. The entries in cell $i,j$ ($i=1,\ldots,10$ and $j=1,\ldots, 10$) are simulated from a Poisson distribution with parameter $ \lambda_{ij }$ defined by 
\begin{equation}
\label{eq:BenchmarkModel}
 \lambda_{ij} = \lambda_0 (1+(i-1) \eta_2) \exp\left(\frac{2(j-1)}{n}\log(
\eta_1)\right),
\end{equation}
where $\lambda_0=$10,000 gives the expected value of claim amounts at the first accident and development year, $\eta_1=0.3$ indicates the reduction of claim amounts at development period $\frac{n}{2}$ compared to the claim amount in the first development year and $\eta_2=0.05$ is the rate of increase for the claim amounts over consecutive accident years. 

\subsubsection{Schiegl method}
\label{sec:SchieglMethod}
This method is proposed in \cite{Schiegl:Handbook} and the simulation of the run-off triangles (containing 10 accident and development years) is based on the collective model, where a different distribution is used for the number of claims and for a single claim amount. The number of claims $N_{ij}$ (for $i=1,\ldots, 10$ and $j=1,\ldots, 10$) follows a Poisson distribution with mean $\lambda_{ij}$ as defined by equation (\ref{eq:BenchmarkModel}) of the benchmark model. Individual claim amounts are then generated using the Gamma distribution with mean $r=$ 2,000 such that the final data are obtained as follows 
\[ Y_{ij} \sim \sum_{k=1}^{N_{
ij }}{Z_k} \text{ with } N_ {ij} \sim \mbox{Poisson}\left(\lambda_{ij}\right) \text{ and } Z_k \sim 
\mbox{ Gamma}\left(r,1\right). \]

\subsubsection{Kaishev method}
\label{sec:KaishevMethod}

\cite{Cowell:Bootstrap} noted that, given the same set of parameter values, the distribution of reserves in the Schiegl method is the same regardless of the sample drawn for the upper triangle (since each $C_{ij}$ value is sampled independently from the others). Therefore, \cite{Cowell:Bootstrap} presents a slightly more complex model (proposed by Kaishev through personal communication) so that, even if all parameters are kept fixed, different simulations will lead to different distributions of reserves.  

As before, the method is based on the collective model and we again generate individual claim amounts using a Gamma distribution with mean $r=$2,000.  The total number of claims $N_i$ for accident year $i=1,\ldots, 10$ is generated from a Poisson distribution with mean $\lambda_i$. 
Each claim (for $k=1,\ldots, N_i)$ follows a Gamma distribution (yielding a certain claim amount $Z_k$) and this amount is then assigned randomly to a certain development year $j=1,\ldots, 10$. This random allocation is done using a $\mbox{Beta}(\alpha=1,\beta=1.6)$ distribution (as suggested in \cite{Cowell:Bootstrap}).  Similar as in the Schiegl method we defined $\lambda_i$ as \[ \lambda_{i} = \lambda_0  (1+(i-1) \eta_2) \] where $\lambda_0$ and $\eta_2$ were also chosen as before. 

Note that plugging in a Pareto or a Lognormal distribution for the individual claim size (instead of a Gamma distribution) in the Schiegl or Kaishev model yielded similar conclusions.

\subsection{Simulation setup}
\label{sec:SimulationSetup}
During the simulation study we consider the following bootstrap procedures: 
\begin{itemize}
\item TCL: standard approach using the classical chain-ladder method and standard bootstrap
\item CB: robust chain-ladder method combined with the classical bootstrap
\item WB: winsorized bootstrap 
\item FRB: fast and robust bootstrap procedure 
\item IFB: influence function bootstrap
\end{itemize}

Note that for IFB, the weight function to determine the resampling probability of each residual contains three tuning parameters, see equation (\ref{weightIFB}). These values need to be chosen to guarantee a sufficiently high efficiency at the uncontaminated case whilst offering robustness when contamination is present. Taking into account this tradeoff between efficiency and robustness, the parameter $c$ is calculated as the $90\%$ quantile of the $\mbox{RESIF}_{ij}$ values, whereas the parameters $d$ and $\gamma$ were set to $30$ and $10$ respectively.  For the winsorized bootstrap the $10\%$ highest and $10\%$ smallest residuals are winsorized. 

The bootstrap procedures are tested for both clean and contaminated data. Two 
contamination settings are shown. Firstly, contamination 
was generated by multiplying observations $Y_{23}$ and $Y_{32}$ in the 
original clean data by a factor of five. Secondly, 
observations $Y_{23}$ and $Y_{41}$ in the original clean data are 
multiplied by two and a half. We also studied other locations and multiplication factors to create outliers, but these results led to similar conclusions. The corresponding tables are not included in the paper, since they do not add new insights in the proposed methodology.

\subsection{Simulation results}
\label{sec:DiscussionOfSimulationResults}

\subsubsection{Benchmark model}
\label{sec:BenkchmarkModel}

For each bootstrap procedure, we investigated the quantiles obtained from 10,000 bootstrap runs. Since we simulate complete data using the benchmark model, the true total reserve can be calculated from the lower triangle according to equation \eqref{eq:TotalReserve}.  Hence, for 10,000 generated data sets we recorded whether its true total reserve is smaller than the calculated bootstrap quantile, for different probability levels. These coverage percentages are shown in Table \ref{tab:Benchmark} for different bootstrap procedures and confidence levels. If the bootstrap procedure is able to approximate the $(1-\alpha)$-quantiles with $\alpha \in (0,0.5)$ well, one may expect the true total reserve to be smaller than the bootstrap quantile in $(1-\alpha)\%$ of the cases.

\begin{table}[ht]
\small
\centering
\begin{minipage}[t]{0.45\textwidth}
\begin{tabular}{crrrrr}
  \toprule
   & & 75\% & 90\% & 95\% & 99.5\% \\ 
  \hline
   &TCL & 71.50 & 86.60 & 92.00 & 98.50 \\ 
	\multirow{3}{*}{\rotatebox{90}{Clean}}& CB & 72.60 & 88.00 & 91.80 & 98.70 \\
	 &WB & 66.50 & 78.60 & 84.80 & 93.60 \\
   &FRB & 69.60 & 83.50 & 89.20 & 96.70 \\
   &IFB & 72.00 & 86.80 & 92.40 & 98.70 \\
	\hline
   &TCL & 100.00 & 100.00 & 100.00 & 100.00 \\ 
  \multirow{3}{*}{\rotatebox{90}{Cont 1}} & CB & 87.00 & 100.00 & 100.00 & 100.00 \\
  &WB & 65.50 & 79.30 & 83.80 & 94.40 \\
    &FRB & 70.00 & 85.70 & 91.30 & 97.90 \\
   &IFB & 63.10 & 80.80 & 88.40 & 97.60 \\
	\hline
   & TCL & 100.00 & 100.00 & 100.00 & 100.00 \\ 
  \multirow{3}{*}{\rotatebox{90}{Cont 2}} & CB & 92.00 & 100.00 & 100.00 & 100.00 \\ 
    &WB  & 74.90 & 85.70 & 90.00 & 96.60 \\
    &FRB & 77.40 & 89.40 & 93.50 & 98.50 \\ 
    &IFB & 76.20 & 91.80 & 98.30 & 100.00 \\ 
   \bottomrule
\end{tabular}
\caption{Coverage percentages for the benchmark model using the Cordeiro adjusted residuals $\boldsymbol{r}^C$ in the bootstrap process.}
\label{tab:Benchmark}
\end{minipage}
\hfill
\begin{minipage}[t]{0.45\textwidth}
\begin{tabular}{crrrrr}
  \toprule
   & & 75\% & 90\% & 95\% & 99.5\% \\ 
  \hline
   &TCL & 67.80 & 81.30 & 87.50 & 96.00 \\ 
	\multirow{3}{*}{\rotatebox{90}{Clean}}& CB & 69.00 & 82.50 & 88.60 & 96.80 \\ 
	&WB & 63.00 & 73.90 & 79.00 & 89.40\\
   & FRB & 67.70 & 80.70 & 86.80 & 95.20 \\ 
    &IFB & 67.90 & 81.50 & 87.60 & 96.40 \\ 
	\hline
   &TCL & 100.00 & 100.00 & 100.00 & 100.00 \\ 
  \multirow{3}{*}{\rotatebox{90}{Cont 1}} & CB & 81.40 & 100.00 & 100.00 & 100.00 \\ 
  &WB & 63.20 & 73.30 & 79.90 & 89.50 \\ 
    &FRB & 68.30 & 83.20 & 89.00 & 96.90 \\ 
   &IFB & 60.80 & 76.00 & 83.20 & 94.70 \\ 
	\hline
   & TCL & 100.00 & 100.00 & 100.00 & 100.00 \\ 
  \multirow{3}{*}{\rotatebox{90}{Cont 2}} & CB & 86.40 & 100.00 & 100.00 & 100.00 \\ 
  &WB & 71.90 & 80.70 & 85.30 & 93.20 \\ 
    &FRB & 75.90 & 87.50 & 91.60 & 97.70 \\ 
   &IFB & 75.50 & 93.60 & 100.00 & 100.00 \\ 
   \bottomrule
\end{tabular}
\caption{Coverage percentages for the benchmark model using the original Pearson residuals $\boldsymbol{r}$ in the bootstrap process.}
\label{tab:benchPure}
\end{minipage}
\end{table}

For the uncontaminated data, it is clear that the WB procedure leads to significantly lower coverage percentages than the other methods.  
The TCL, CB and IFB methodologies give comparable results whereas the FRB procedure yields slightly lower coverage probabilities (but still much better than the WB method). 

To study the effect of outliers, we repeated the simulation study using the first (i.e. multiplying $Y_{23}$ and $Y_{32}$ by five) and second (i.e. multiplying $Y_{23}$ and $Y_{41}$ by two and a half) contamination setting. In both situations, the FRB and IFB clearly outperform the other bootstrap procedures. We immediately see that the TCL and CB procedures are highly 
influenced by the outliers as almost all the coverage percentages equal $100\%$ (this is because the obtained bootstrap quantiles became very large due to the outliers), whereas the WB bootstrap quantiles are too small leading to the smallest coverage percentages. It can be seen that in contamination scenario two the estimates of the $95\%$ and $99.5\%$ quantiles obtained by the IFB bootstrap procedure are still affected by the outliers. In this setting the outliers are closer to the real data and therefore it is more difficult to detect them. 

To study the effect of the residual adjustment on the coverage percentages, Table \ref{tab:benchPure} shows the percentages obtained by using the original Pearson residuals. These coverage percentages are lower and further away from the target than those obtained with the adjusted Pearson residuals. 
The coverage percentages based on the adjusted residuals using the \cite{England:Analytic} adjustment factor ($\boldsymbol{r}^E$) and the correction proposed by \cite{Pinheiro:BootstrapClaims} ($\boldsymbol{r}^P$) were comparable to the results found in Table \ref{tab:Benchmark} (using $\boldsymbol{r}^C$) and are therefore omitted. 

To compare the speed of the different bootstrap procedures, the average 
computation time for a bootstrap run consisting of 10,000 bootstrap samples 
for a run-off triangle, simulated using the benchmark model, is shown in Table 
\ref{tab:ComputationTimes}. These computations were performed on an
Intel Ivy Bridge Xeon E5-2680V2 2,8 GHz CPU and averaged across 100 runs. 

\begin{table}[htbp]
	\centering
		\begin{tabular}{l|ccccc}
		\toprule
						 & TCL & CB & WB & IFB & FRB\\ \midrule
		mean     & 0.98 & 6.75 & 19.47 & 1.03 & 0.016\\
		st. dev. & 0.06 & 2.90 & 14.44 & 0.05 & 0.002\\
		\bottomrule
		\end{tabular}
	\caption{Computation times expressed in minutes.}
	\label{tab:ComputationTimes}
\end{table}

The FRB method clearly outperforms the other alternatives in terms of computational efficiency. Given the high time complexity and bad performance of the CB and WB
methods in the (contaminated) benchmark model, we discard these procedures in further studies. The other procedures are now compared using the more advanced simulation models. 

\FloatBarrier

\subsubsection{Schiegl and Kaishev model}
\label{sec:SchieglAndKaishevModel}

\begin{figure}[ht!]
\centering
\begin{minipage}[l]{0.45\textwidth}
	\centering
	\includegraphics[trim=5cm 0cm 3cm 0cm, width=1\textwidth]{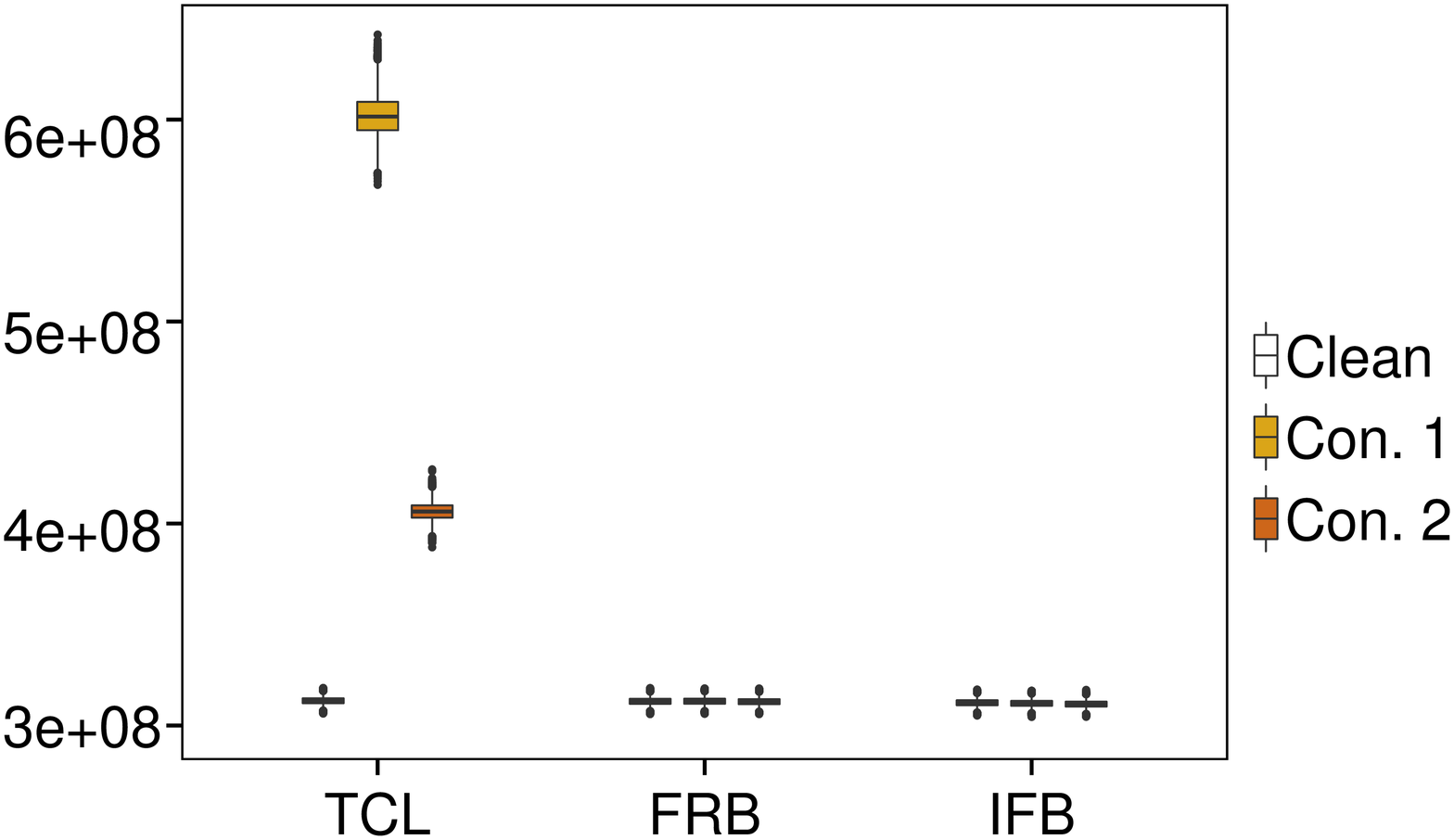}
	\caption{Boxplots of the obtained $99.5\%$ quantiles of the 10,000 
bootstrapped estimates using Schiegl model.}
	\label{fig:Schiegl}
\end{minipage}
\hfill
\begin{minipage}[l]{0.45\textwidth}
	\centering
		\includegraphics[trim=6cm 1cm 3cm 0cm, width=1\textwidth]{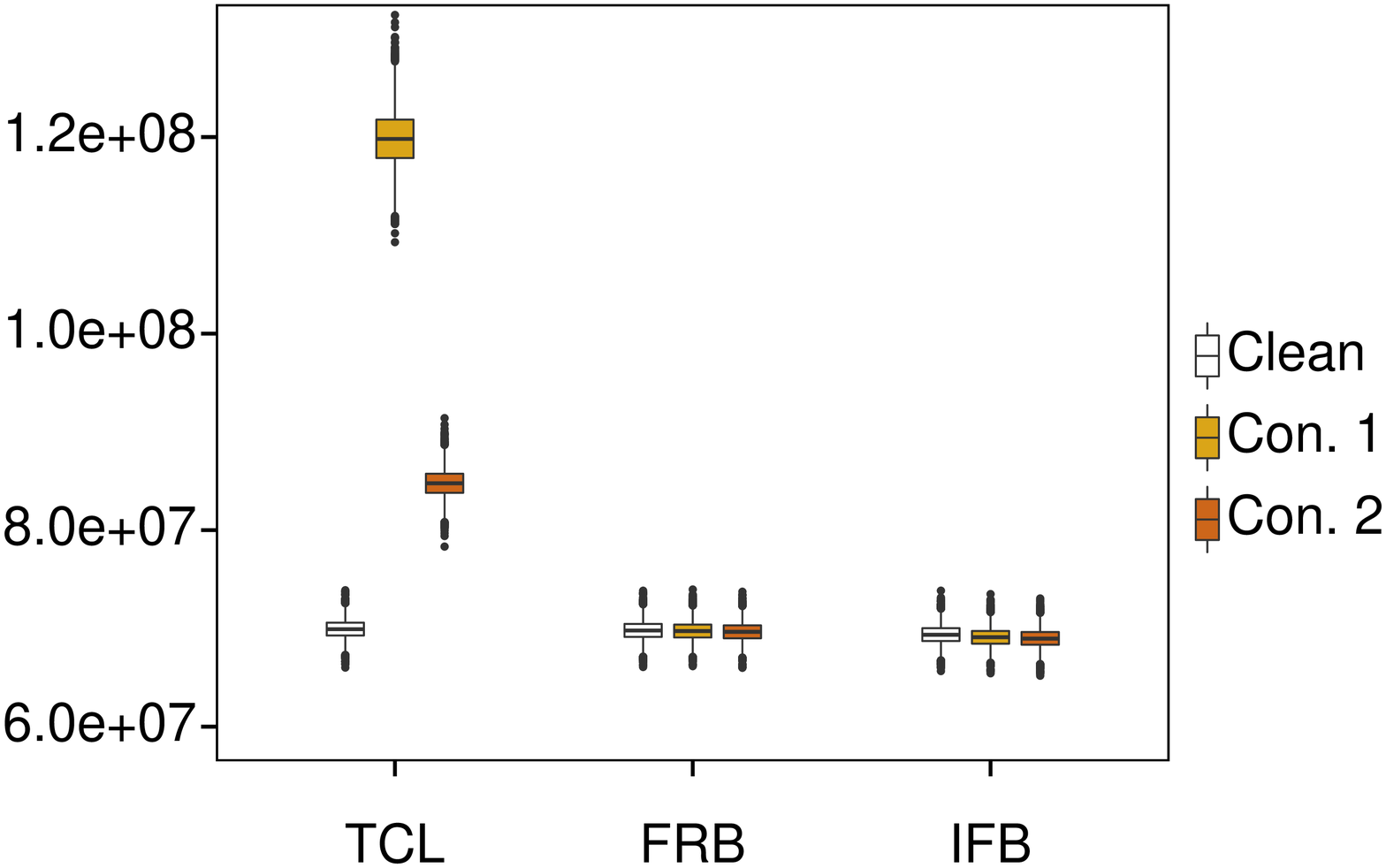}
	\caption{Boxplots of the obtained $99.5\%$ quantiles of the 10,000 
bootstrapped estimates using Kaishev model.}
	\label{fig:Kaishev}
\end{minipage}
\end{figure}

We have applied the remaining bootstrap techniques on clean and contaminated data sets simulated by the Schiegl and Kaishev models. Boxplots of the obtained $99.5\%$ quantiles based on 10,000 bootstrapped estimates for the 
total reserve are shown in Figures \ref{fig:Schiegl} and \ref{fig:Kaishev}. The boxes in the boxplots are white for the clean data, orange for the first contamination setting (i.e. multiplying $y_{23}$ and $y_{32 }$ by 5) and red for the second contamination setting (i.e. multiplying $y_{23}$ and $y_{41}$ by 2.5).  

\begin{figure}[ht!]
\centering
\begin{minipage}[l]{0.45\textwidth}
	\centering
		\includegraphics[trim=7cm 0cm 3cm 0cm, width=1\textwidth]{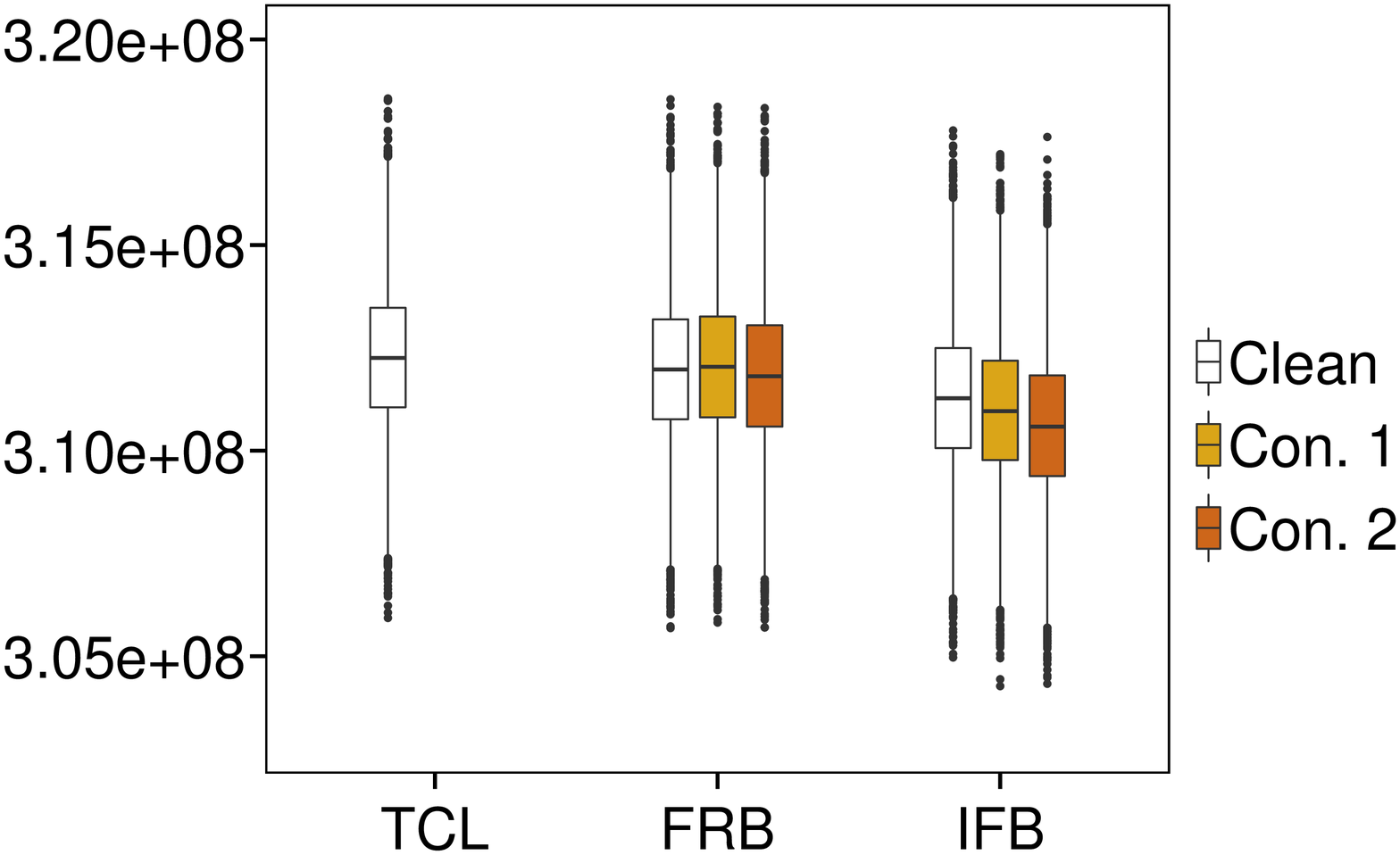}
	\caption{Detailed version of Figure \ref{fig:Schiegl} (discarding large values on $y$-axis).}
	\label{fig:SchieglZoom}
\end{minipage}
\hfill
\begin{minipage}[l]{0.45\textwidth}
	\centering
		\includegraphics[trim=7cm 0cm 3cm 0cm, width=1\textwidth]{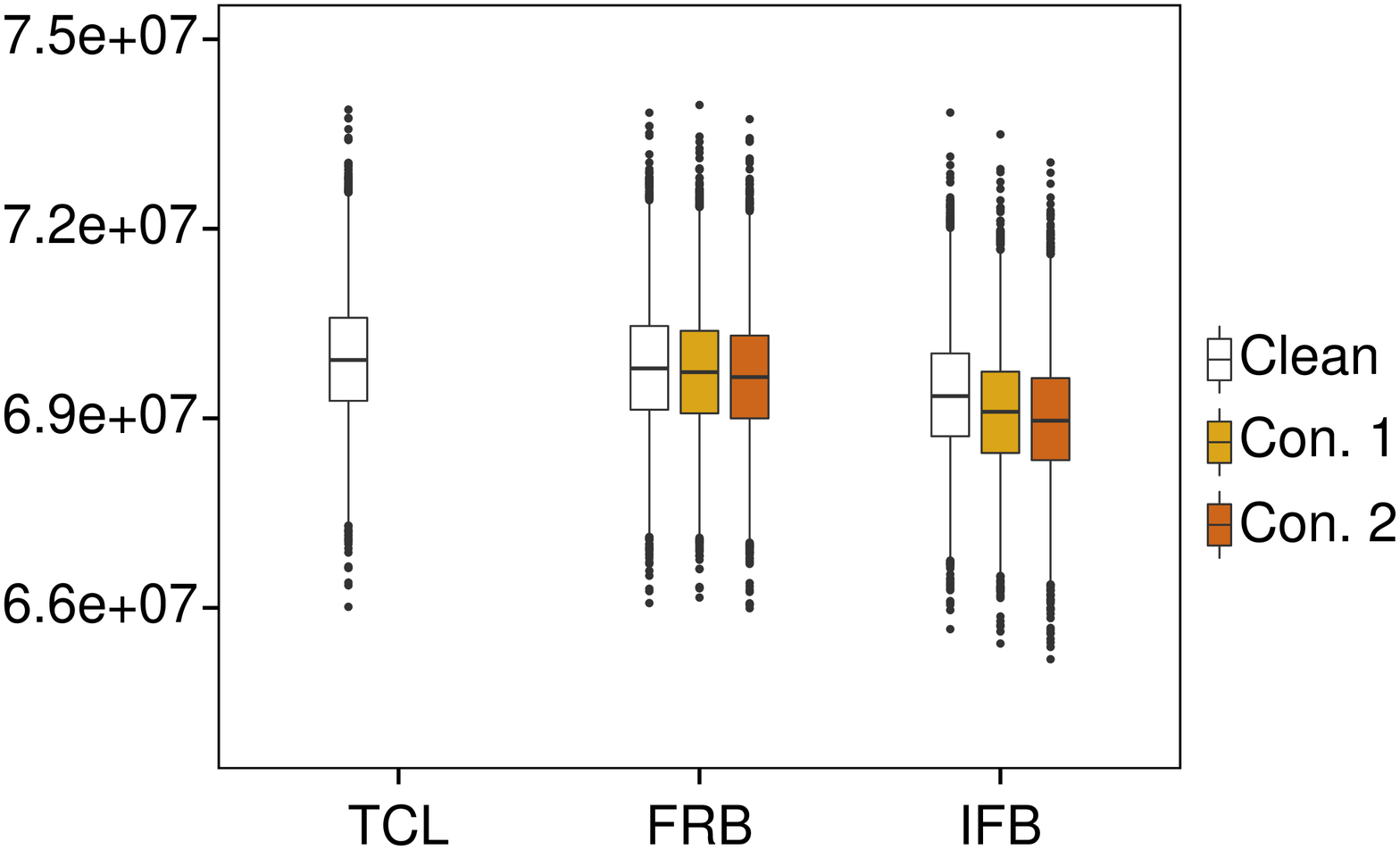}
	\caption{Detailed version of Figure \ref{fig:Kaishev} (discarding large values on $y$-axis).}
	\label{fig:KaishevZoom}
\end{minipage}
\end{figure}

It is immediately clear from Figures \ref{fig:Schiegl} and \ref{fig:Kaishev} 
that TCL breaks down as soon as the data is contaminated, whereas the other 
methods give results that are almost the same as those obtained by the 
classical method on the initial clean data without outliers. The presence of the outliers thus only has a minor effect on these methods.
In order to better compare the robustness properties of IFB and FRB, we have removed the results 
for TCL on contaminated data in Figures \ref{fig:SchieglZoom} and 
\ref{fig:KaishevZoom}. When no outliers are included, the boxplots for FRB are nearly identical to the ones obtained with TCL on the clean data, whereas the results for IFB are clearly lower. These differences are even more obvious when looking at the contaminated settings: the boxplots for FRB hardly change when outliers are added, whereas the boxplots for IFB are getting further away from the target (i.e. the results obtained with TCL on the clean data). 

The same conclusions are obtained from the results in Tables \ref{tab:Schiegl} and \ref{tab:Kaishev}, where the 10,000 bootstrapped estimates for the total reserve are compared with the true reserve (obtained by summing up the lower triangle of the simulated data set). 
The percentages in the tables again represent in how many of the 10,000 runs the true reserve was covered by the corresponding confidence interval for different confidence levels. 
For the uncontaminated data the coverage percentages for TCL are fairly 
close to their expected values with FRB percentages being only slightly 
smaller. The IFB method achieves smaller coverage percentages. When outliers are added, TCL clearly breaks down: the bootstrap quantiles become very large, which explains the high coverage rates in almost all situations. However, for data simulated from the Schiegl model and the second contamination setting, the $75\%$ bootstrap quantiles never succeed in covering the real future reserve. Similar observations can be made for the Kaishev model. 
On the other hand, the coverage percentages 
for the robust methods always remain close to the target percentages. The results for FRB are again more stable than for IFB. 

\begin{table}[ht]
\small
\centering
\begin{minipage}[t]{0.45\textwidth}
\begin{tabular}{crrrrr}
  \toprule
   & & 75\% & 90\% & 95\% & 99.5\% \\ 
  \hline
  \multirow{3}{*}{\rotatebox{90}{Clean}}  &TCL & 71.84 & 87.09 & 92.66 & 98.62 \\ 
   & FRB & 71.02 & 85.75 & 91.34 & 97.90 \\
    &IFB & 67.13 & 80.50 & 86.67 & 95.51 \\ 
	\hline
   \multirow{3}{*}{\rotatebox{90}{Cont 1}}&TCL & 100.00 & 100.00 & 100.00 & 100.00 \\ 
    &FRB & 65.37 & 83.18 & 90.31 & 98.14 \\ 
   &IFB & 55.43 & 72.31 & 80.89 & 93.79 \\
	\hline
   \multirow{3}{*}{\rotatebox{90}{Cont 2}} &TCL & 0.00 & 100.00 & 100.00 & 100.00 \\
    &FRB & 59.23 & 79.12 & 87.47 & 97.35 \\ 
   &IFB & 47.68 & 65.64 & 74.91 & 90.89 \\ 
   \bottomrule
\end{tabular}
\caption{Coverage results for Schiegl model.}
\label{tab:Schiegl}
\end{minipage}
\hfill
\begin{minipage}[t]{0.45\textwidth}
\begin{tabular}{crrrrrr}
  \toprule
 && 75\% & 90\% & 95\% & 99.5\% \\ 
  \hline
  \multirow{3}{*}{\rotatebox{90}{Clean}}&TCL & 72.95 & 88.02 & 93.01 & 98.75 \\ 
   &FRB & 72.64 & 87.06 & 91.84 & 98.21 \\  
  &IFB & 67.52 & 80.86 & 86.93 & 95.23 \\  
	\hline
  \multirow{3}{*}{\rotatebox{90}{Cont 1}}&TCL & 2.99 & 100.00 & 100.00 & 100.00 \\ 
   &FRB & 62.71 & 81.98 & 89.41 & 97.97 \\ 
  &IFB & 52.06 & 69.43 & 78.40 & 92.50 \\ 
	\hline
  \multirow{3}{*}{\rotatebox{90}{Cont 2}}&TCL & 0.01 & 97.59 & 100.00 & 100.00 \\
   &FRB  &  58.70 & 79.41 & 87.80 & 97.55 \\ 
   &IFB &  46.67 & 65.20 & 74.63 & 90.77 \\
   \bottomrule
\end{tabular}
\caption{Coverage results for Kaishev model.}
\label{tab:Kaishev}
\end{minipage}
\end{table}

\subsubsection{Conclusion}
\label{sec:Conclusion}
From this simulation study we can conclude that TCL and CB are very sensitive to outliers and that WB consistently underestimates the total reserve quantile. The IFB and FRB procedures yield efficient and robust results and FRB is generally less influenced by outliers than IFB. Taking computational speed into account as well, we advice to combine the robust chain-ladder method with the FRB approach for prediction purposes in the claims reserving framework.

\section{Real data application }
\label{sec:RealDataApplication}

We now apply our optimal robust methodology (i.e. robust chain-ladder method combined with the FRB approach) to a run-off triangle from practice, which is shown in Table \ref{table:ScheduleP}. 

 \begin{table}[!ht]
\scriptsize
\begin{center}
\begin{tabular}{c|r r r r r r r r r r}    
\toprule
       &1   &2   &3   &4   &5   &6   &7  &8   &9  &10\\ \hline
  1988 &\cellcolor{gray!40}794 &\cellcolor{gray!40}569 &\cellcolor{gray!40}497 &\cellcolor{gray!40}409 &\cellcolor{gray!40}221 &\cellcolor{gray!40} 55 &\cellcolor{gray!40} 62&\cellcolor{gray!40}  9 &\cellcolor{gray!40} 0 &\cellcolor{gray!40} 0\\
  1989 &\cellcolor{gray!40}696 &\cellcolor{gray!40}499 &\cellcolor{gray!40}429 &\cellcolor{gray!40}437 &\cellcolor{gray!40}227 &\cellcolor{gray!40}154 &\cellcolor{gray!40} 10&\cellcolor{gray!40} 57 &\cellcolor{gray!40} 0 & 0\\
  1990 &\cellcolor{gray!40}799 &\cellcolor{gray!40}690 &\cellcolor{gray!40}391 &\cellcolor{gray!40}529 &\cellcolor{gray!40}326 &\cellcolor{gray!40} 34 &\cellcolor{gray!40}  4&\cellcolor{gray!40}37 &45 & 0\\
  1991 &\cellcolor{gray!40}796 &\cellcolor{gray!40}405 &\cellcolor{gray!40}234 &\cellcolor{gray!40}287 &\cellcolor{gray!40}131 &\cellcolor{gray!40}263 &\cellcolor{gray!40}438&  0 & 1 & 4\\
  1992 &\cellcolor{gray!40}742 &\cellcolor{gray!40}498 &\cellcolor{gray!40}368 &\cellcolor{gray!40}199 &\cellcolor{gray!40}346 &\cellcolor{gray!40}49 &  1&  0 & 0 & 1\\
  1993 &\cellcolor{gray!40}637 &\cellcolor{gray!40}518 &\cellcolor{gray!40}140 &\cellcolor{gray!40}231 &\cellcolor{gray!40}79 & 34 & 19& 89 & 5 & 0\\
  1994 &\cellcolor{gray!40}703 &\cellcolor{gray!40}427 &\cellcolor{gray!40}364 &\cellcolor{gray!40}194 & 98 & 40 & 15&  6 &61 & 3\\
  1995 &\cellcolor{gray!40}580 &\cellcolor{gray!40}375 &\cellcolor{gray!40}227 &132 &121 & 18 & 10& 52 &45 & 6\\
  1996 &\cellcolor{gray!40}563 &\cellcolor{gray!40}266 &106 &203 & 71 & 30 & 23&  4 & 3 &19\\
  1997 &\cellcolor{gray!40}324 &193 &173 & 30 & 36 & 49 & 44& 19 & 3 & 0\\
	\bottomrule
\end{tabular}
\end{center}
\caption{\label{table:ScheduleP} Real run-off triangle from Schedule P data.}
\end{table}

This real data set is extracted from the schedule P 
- Analysis of Losses and Loss Expenses in the National Association 
of Insurance Commissioners (NAIC) database and is available on 
the website of the Casualty Actuarial Society \footnote{http://www.casact.org/research/index.cfm?fa=loss_reserves_data}. 
This database contains data on several lines of business for a variety 
of businesses in the United States and here we focus on the
\textit{other liability line of business} for the Rockford Mutual Insurance Company. The true total reserve is 1812 (the sum of the observations in the lower triangle of Table \ref{table:ScheduleP}). 

\begin{figure}
	\centering
		\includegraphics[width=0.50\textwidth]{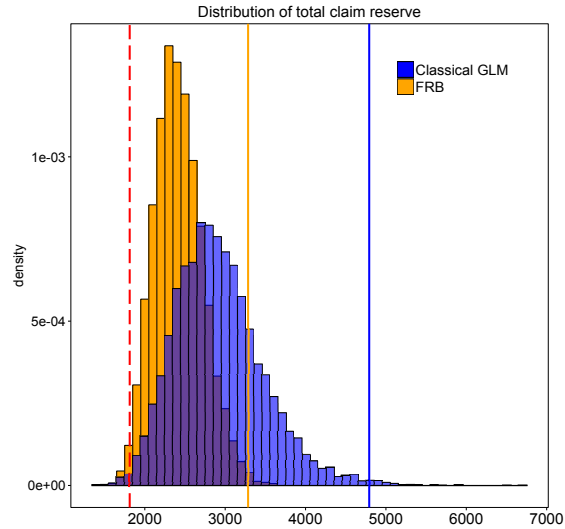}
	\caption{Bootstrapped overal reserve estimate for the Rockford Mutual Insurance company.}
	\label{fig:ScheduleP_Distribution_TotalReserve}
\end{figure}

We recommend to always apply both the classical and robust procedures to a given dataset and compare their results. For this example, the classical chain-ladder method leads to an total reserve estimate of 2837 whereas the robust alternative leads to a total claim reserve estimate of 2304, which is a difference of roughly 20$\%$. The obtained bootstrap $99.5\%$ quantile for the total reserve is 4781 for the classical methodology and 3285 for the robust procedure based on FRB. This clear discrepancy may indicate the presence of possible outlier(s) in the data. The bootstrap distributions are shown in figure \ref{fig:ScheduleP_Distribution_TotalReserve}. The true total reserve is marked with a red dashed line. 
		
An advantage of the robust chain-ladder method is that the most influential points can easily be detected using	
the weights of the robust chain-ladder method defined in equation (\ref{eq:robweights}). From Table \ref{tbl:GraphRep_Weights} it is clear that the sixth and seventh development year of accident year 1991 have a very low 
weight and should be examined in more detail. If these observations are actually errors, then the robust estimate is a reliable total reserve estimate, otherwise the total reserve estimate should be adjusted accordingly (taking into account the possibility of observing such atypical claims in future). 

\begin{table}[!ht]
\scriptsize
\centering
	\begin{tabular}{l|rrrrrrrrrr}
  \hline
 & 1 & 2 & 3 & 4 & 5 & 6 & 7 & 8 & 9 & 10 \\ 
  \hline
  1988 & 1.00 & 1.00 & 1.00 & 1.00 & 1.00 & 1.00 & 1.00 & 1.00 & 1.00 & 1.00 \\ 
  1989 & 1.00 & 1.00 & 1.00 & 1.00 & 1.00 & \cellcolor{red!37}0.63 & 1.00 & 1.00 & 1.00 &  \\ 
  1990 & 1.00 & 1.00 & 1.00 & \cellcolor{red!07}0.93 & 1.00 & 1.00 & 1.00 & 1.00 &  &  \\ 
  1991 & 1.00 & 1.00 & 1.00 & 1.00 & 1.00 & \cellcolor{red!78}0.22 & \cellcolor{red!93}0.07 &  &  &  \\ 
  1992 & 1.00 & 1.00 & 1.00 & \cellcolor{red!17}0.83 & \cellcolor{red!43}0.57 & 1.00 &  &  &  &  \\ 
  1993 & 1.00 & \cellcolor{red!20}0.80 & \cellcolor{red!24}0.76 & 1.00 & \cellcolor{red!09}0.91 &  &  &  &  &  \\ 
  1994 & 1.00 & 1.00 & 1.00 & 1.00 &  &  &  &  &  &  \\ 
  1995 & 1.00 & 1.00 & 1.00 &  &  &  &  &  &  &  \\ 
  1996 & 1.00 & 1.00 &  &  &  &  &  &  &  &  \\ 
  1997 & 1.00 &  &  &  &  &  &  &  &  &  \\ 
   \hline
\end{tabular}
  \caption{Weights from the robust chain-ladder method for the Schedule P data.}%
  \label{tbl:GraphRep_Weights}%
\end{table}

\begin{table}[!ht]
\scriptsize
\centering
\begin{tabular}{r|rrrrrrrrrr}
  \hline
 & 1 & 2 & 3 & 4 & 5 & 6 & 7 & 8 & 9 & 10 \\ 
  \hline
  1988 & -0.94 & 0.70 & 6.35 & 2.76 & -1.06 & -5.44 & -5.82 & -4.28 & -0.02 & 0.00 \\ 
  1989 & -3.25 & -1.34 & 3.71 & 5.13 & -0.04 & 4.40 & -10.17 & 4.28 & 0.02 &  \\ 
  1990 & -2.78 & 3.96 & -0.54 & 7.45 & 4.45 & -7.92 & -11.37 & 0.09 &  &  \\ 
  1991 & -0.56 & -6.05 & -7.07 & -3.51 & -6.77 & 14.33 & 27.72 &  &  &  \\ 
  1992 & 0.20 & 0.10 & 1.77 & -6.77 & 9.12 & -5.19 &  &  &  &  \\ 
  1993 & 3.13 & 7.16 & -7.27 & -0.87 & -6.56 &  &  &  &  &  \\ 
  1994 & 1.73 & -0.77 & 3.69 & -5.45 &  &  &  &  &  &  \\ 
  1995 & 1.05 & 0.07 & -1.64 &  &  &  &  &  &  &  \\ 
  1996 & 3.03 & -3.70 &  &  &  &  &  &  &  &  \\ 
  1997 & -0.00 &  &  &  &  &  &  &  &  &  \\ 
   \hline
\end{tabular}
  \caption{Pearson residuals obtained from the classical chain-ladder algorithm for the Schedule P data.}%
  \label{tbl:GraphRep_PRes}%
\end{table}

For comparison, the Pearson residuals obtained using the classical chain-ladder method are shown in Table \ref{tbl:GraphRep_PRes}. Development year six and seven of accident year 1991 have the largest residuals, but other observations also have large residuals (e.g. above 5 or even above 10 in absolute value).  
It is thus not obvious to determine how many observations have an unusually large residual and it is also not  clear what adjustments should be made to these observations or how they should be treated in the bootstrap process 
On the other hand, our robust techniques yield an automatic, data driven and objective methodology to treat and detect possible outliers.

Under the new Solvency II regulations insurers are also compelled to investigate 
one year risk measures \citep{BAJ:SolvencyII}. \cite{Wutrich:CDR} proposed to 
use the Claims Development Result (\mbox{CDR}) for this purpose. Let $D_k = \left\{Y_{ij} \, | 
\, i + j \leqslant k + 1 \right\}$ denote the upper triangle of stochastic variables up to 
calender year k with $n\leqslant k \leqslant 2n-1$. Note that for $k=n$ this corresponds to the past claims data and for $k=2n-1$ the full square of data is available. The true \CDR for claims in accident year $i \in \left\{1,
\ldots,n\right\}$ during the next calender year $(k, k+1]$ is defined as \[ 
\mbox{CDR}_i(k+1) = \mbox{E}\left[ \mbox{TR}_i \,|\, D_k \right] - (Y_{i,k-i +
 1}  + \mbox{E}\left[ \mbox{TR}_i \,|\, D_{k+1} \right]),\] where $\mbox{TR}_i = \sum_{j=k-i+1}^{n}{Y_{ij}}
$ denotes the total claims amount for accident year $i$. The aggregated \CDR 
is then defined as \[ \mbox{CDR}(k+1) = \sum_{i=1}^{n}{\mbox{CDR}_i(k+1)}. \]

To estimate the \CDR the stochastic quantities are replaced by their observations and 
estimates respectively. Several authors already considered this problem including 
\cite{Diers:StochReReserving} and \cite{Ohlsson:OneYear}. \cite{Lacoume:OneYear}, 
\cite{Boisseau:SolvII} and \cite{Boumezoued:OneYear} complement these papers 
and propose to add the following steps to the bootstrap procedure (explained in Section \ref{sec:Bootstrapping}) in every run $b=1,\ldots, B$:

\begin{enumerate}
	\item[4.] Denote the fitted values obtained from the GLM in step 3 of the bootstrap loop as $\hat{Y}^{(b)}_{ij}$, see Table \ref{tbl:GraphRep_CDR}(b). The bootstrap reserve $R^{(b)}$ then corresponds to $\sum_{i=2}^{n}{\sum_{j=n-i+2}^{n}{\hat{Y}^{(b)}_{ij}}}$. An extended bootstrap history is constructed by adding the first fitted anti-diagonal $\hat{Y}^{(b)}_{ij}$ with $i+j=n+2$ to the bootstrapped historic observations. Based on this extended bootstrap history the GLM is refitted leading to the fitted values $\widetilde{Y}^{(b)}_{ij}$, see Table \ref{tbl:GraphRep_CDR}(c). This leads to a new total future reserve estimate $R^{(b)}_{ext} = \sum_{i=3}^{n}{\sum_{j=n-i+3}^{n}{\widetilde{Y}^{(b)}_{ij}}}$.
	\item[5.] A bootstrapped value for the \CDR may then be obtained by computing : 
					 \[\mbox{CDR}^{(b)}(k+1) = R^{(b)} - R^{(b)}_{ext} = \sum_{i=2}^{n}{\sum_{j=n-i+2}^{n}{\hat{Y}^{(b)}_{ij}}} - \sum_{i=3}^{n}{\sum_{j=n-i+3}^{n}{\widetilde{Y}^{(b)}_{ij}}}.  \]
\end{enumerate}

\begin{figure}[!htb]
        \centering
				\includegraphics[width=0.5\textwidth]{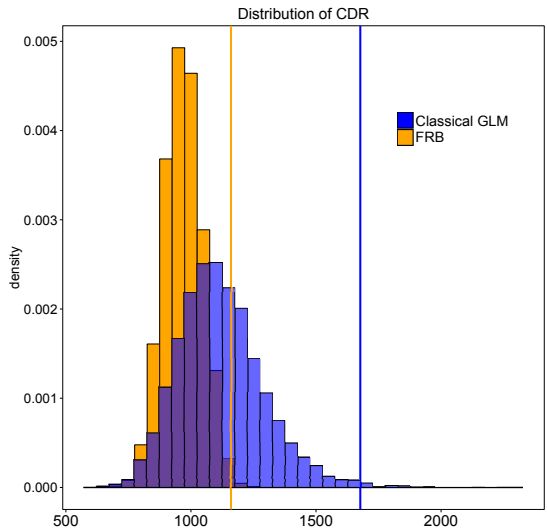}
				\caption{Estimated distribution of the CDR for the Rockford Mutual Insurance Company using the classical GLM approach in blue (dark color) and the FRB approach in orange (light color).}
				\label{fig:DistributionOneYearTotal}
\end{figure}

The bootstrapped \CDR is shown in Figure \ref{fig:DistributionOneYearTotal}. 
The $99.5\%$ quantile corresponds to 1,677 for the classical procedure 
and 1,160 for the robust procedure. As the Value at Risk under the 
Solvency II regulations is  directly related to this quantile, this example 
again shows the impact of outliers on standard claims reserving results and the good performance of the proposed robust alternative.

\begin{table}[!ht]
\tiny
\centering
\centering
  \subfloat[][Obtain a bootstrap triangle]{
	\begin{tabular}{c|c c c c}
\centering
&1&2&\ldots&$J$\\
\hline
1&$\cellcolor{gray!40}Y_{11}^{(b)}$&\cellcolor{gray!40}$Y_{12}^{(b)}$&\cellcolor{gray!40}$\ldots$&\cellcolor{gray!40}$Y_{1J}^{(b)}$\\
2&$\cellcolor{gray!40}Y_{21}^{(b)}$&\cellcolor{gray!40}$Y_{22}^{(b)}$&\cellcolor{gray!40}$\ldots$& \\
$\ldots$&\cellcolor{gray!40}$\ldots$&\cellcolor{gray!40}$\ldots$&\\
$I$&\cellcolor{gray!40}$Y_{I1}^{(b)}$& & & \\
\end{tabular}
}%
  \qquad
	\hfill
	\subfloat[][Fit a GLM to the bootstrap history completing the triangle.]{
\begin{tabular}{c|c c c c}
\centering
&1&2&\ldots&$J$\\
\hline
1&\cellcolor{gray!40}$Y_{11}^{(b)}$&\cellcolor{gray!40}$Y_{12}^{(b)}$&\cellcolor{gray!40}$\ldots$&\cellcolor{gray!40}$Y_{1J}^{(b)}$\\
2&\cellcolor{gray!40}$Y_{21}^{(b)}$&\cellcolor{gray!40}$Y_{22}^{(b)}$&\cellcolor{gray!40}$\ldots$&$\hat{Y}_{2J}^{(b)}$\\
$\ldots$&\cellcolor{gray!40}$\ldots$&\cellcolor{gray!40}$\ldots$&$\ldots$&$\ldots$\\
$I$&\cellcolor{gray!40}$Y_{I1}^{(b)}$&$\hat{Y}_{I2}^{(b)}$&$\ldots$&$\hat{Y}_{IJ}^{(b)}$\\
\end{tabular}
}
	\hfill
	\subfloat[][Add the first fitted anti-diagonal to the bootstrap history and refit the GLM.]{
	\begin{tabular}{c|c c c c}
\centering
&1&2&\ldots&$J$\\
\hline
1&\cellcolor{gray!40}$Y_{11}^{(b)}$&\cellcolor{gray!40}$Y_{12}^{(b)}$&\cellcolor{gray!40}$\ldots$&\cellcolor{gray!40}$Y_{1J}^{(b)}$\\
2&\cellcolor{gray!40}$Y_{21}^{(b)}$&\cellcolor{gray!40}$Y_{22}^{(b)}$&\cellcolor{gray!40}$\ldots$&\cellcolor{gray!20}$\hat{Y}_{2J}^{(b)}$\\
$\ldots$&\cellcolor{gray!40}$\ldots$&\cellcolor{gray!40}$\ldots$&\cellcolor{gray!20}$\ldots$&$\ldots$\\
$I$&\cellcolor{gray!40}$Y_{I1}^{(b)}$&\cellcolor{gray!20}$\hat{Y}_{I2}^{(b)}$&$\ldots$&$\widetilde{Y}_{IJ}^{(b)}$\\
\end{tabular}
	}
  \caption{Graphical representation of the CDR calculation for bootstrap loop $b$.}%
  \label{tbl:GraphRep_CDR}%
\end{table}

\section{Conclusions}
\label{sec:Conclusions}

The bootstrap procedure is a powerful and popular technique to obtain information from a single sample of data,
which would usually be obtained using analytic techniques. It complements sample estimates with measures of accuracy (e.g. standard error, confidence interval, prediction error). In claims reserving, the bootstrap procedure is typically used to produce a standard error of prediction and an approximate predictive distribution for the random future losses. In this paper we studied the effect of outliers on these procedures and it is shown that it is not sufficient to plug a robust estimator into the classical bootstrap procedure. Therefore, we implemented and compared all existing (up to our knowledge) robust bootstrap methods that could be adapted to the claims reserving framework. The Fast and Robust Bootstrap (FRB) approach appears to be a good choice in the sense that it yields very reliable results and is computationally fast.

The presence of outliers may reveal that the data are more heterogeneous than assumed and cannot be handled by the original statistical model. Outliers can be isolated or may come in clusters, indicating that there are subgroups in the population that behave differently. We believe that a robust analysis together with some exploratory data analysis techniques can reveal structures in the data that would remain hidden in a traditional analysis. Robust methods try to fit the model followed by the majority of the data: if the data contain no outliers, they give approximately the same results as the classical method, while if the data is contaminated they give approximately the same results as the classical method applied to the outlier-free data. The best estimates obtained by the robust methodology is hence the result that one would obtain when applying the traditional methodology on the data without outliers. Note that we do not know the proportion of outliers in advance.

Using our robust methodology, insurers obtain estimates that are similar to the results that  would have been obtained from outlier-free data. Moreover, the procedure also detects outlying observations (if any). Of course, it is important to examine the detected outliers and to understand the reasons for their atypical behaviour. 
In this way the proposed robust methodology is helpful to gain insight in the data. 
If the outliers are errors or due to causes that are not expected to re-occur in the future, then the robust estimates allow to build up more realistic reserve estimates and confidence intervals. 
However, if such atypical observations are likely to happen again in future, it is necessary to model also their process (outside the scope of this paper) and to predict how much extra reserve (besides the robust total reserve estimate) is needed to cope with such outlying observations in future years. In such a case, the final estimate may for instance be equal to the robust total reserve estimate plus a safe margin for the outliers. 

\section*{Acknowledgments}
This work was supported by the Internal Funds KU Leuven under Grants $C16/15/068$ and $C24/15/001$; the Flemish Science Foundation (FWO) under Grant $1523915N$. The computational resources and services used in this work were provided by the VSC (Flemish Supercomputer Center), funded by the Research Foundation - Flanders (FWO) and the Flemish Government – department EWI.

\bibliographystyle{apalike} 
\bibliography{Robustness,RobustnessExtension} 

\section{Appendix}
\label{sec:Appendix}

\subsection{Hat matrix of the robust GLM estimator}
\label{sec:hatRobGLM}
To calculate the adjusted Pearson residuals $\boldsymbol{r}^C$ an expression for the hat matrix corresponding to the robust estimator of \cite{Cantoni:genreg} is needed. To this end we rewrite the estimator as an iteratively weighted least squares problem. Using a first order Taylor expansion of equation (\ref{eq:CantoniPoisGLM}), we obtain 
\begin{equation} 
\label{eq:Cantoni_NewtonRapson}
\thetab^* = \left[ -\frac{\partial}{\partial\thetab}\boldsymbol{\overline{\psi}}(\boldsymbol{y}, \boldsymbol{\mu}) \right]^{-1}\boldsymbol{\overline{\psi}}(\boldsymbol{y}, \boldsymbol{\mu}) + \thetab
\end{equation}
where $ \boldsymbol{\overline{\psi}}(\boldsymbol{y}, \boldsymbol{\mu}) = \sum_{i=1}^n\sum_{j=1}^{n-i+1}{\boldsymbol{\psi}(y_{ij}, \mu_{ij})} = X^t D 1_N$ with $1_N$ a unit vector of length $N$ and D the diagonal matrix with elements  \[\psi_c(r_{ij})w(x_{ij})\frac{1}{V^{1/2}(\mu_{ij})}\mu_{ij} - \frac{2}{n(n+1)}
\sum_{i=1}^n\sum_{j=1}^{n-i+1} E \left[\psi_c(r_{ij})\right]\omega(x_{ij})
\frac{1}{V^{1/2}(\mu_{ij})}\mu_{ij}.\]  
We then replace $-\frac{\partial}{\partial\thetab}\boldsymbol{\overline{\psi}}(\boldsymbol{y}, \boldsymbol{\mu}) $ by $-\mbox{E}\left[-\frac{\partial}{\partial\thetab}\boldsymbol{\overline{\psi}}(\boldsymbol{y}, \boldsymbol{\mu}) \right]$. \cite{Cantoni:genreg} computed this to be equal to \[ X^tBX \] where $B$ is the diagonal matrix with elements $\mbox{E}\left[\psi_c(r_{ij}) \frac{Y_{ij}-\mu_{ij}}{\mu_{ij}} \right]\frac{w(\boldsymbol{x}_{ij})}{\mu^{3/2}_{ij}}$ and
 
\begin{align*}
\mbox{E}\left[\psi_c(r_{ij}) \frac{Y_{ij}-\mu_{ij}}{\mu_{ij}} \right]
= &c\left(\mbox{P}\left(Y_{ij} = j_1\right) + \mbox{P}\left(Y_{ij} = j_2\right) \right) \\
&+\sqrt{\mu_{ij}}\left(\mbox{P}\left(Y_{ij} = j_1-1\right)  - \mbox{P}\left(Y_{ij} = j_1\right) - \mbox{P}\left(Y_{ij} = j_2 - 1 \right) + \mbox{P}\left(Y_{ij} = j_2\right)\right) \\
&+\mu_{ij} \mbox{P}\left(j_1 \leqslant Y_{ij} \leqslant j_2-1\right). 
\end{align*}

Equation (\ref{eq:Cantoni_NewtonRapson}) then reduces to 
\[ \thetab^* =  (X^tBX)^{-1}X^t D 1_N  + \thetab \]
which can be rewritten as 
\[ \thetab^* = (X^tBX)^{-1}X^tB \left[B^{-1}D1_N+ X\thetab  \right], \] assuming $B$ and $(X^tBX)$ are invertible. This expression may be recognized as that of a weighted least squares regression where $\left[B^{-1}D1_N+ X\thetab   \right]$ is regressed onto $X$ with weight matrix $B$. The hat matrix for the robust estimator of \cite{Cantoni:genreg} is thus \[H = X(X^tBX)^{-1}X^tB.\] 

\subsection{Calculation of the correction terms proposed by Cordeiro (2004)}
\label{sec:Cordeiro}
Based on the general expression given by \cite{Cordeiro_PResidGLM}, the following vector of expected Pearson residuals is obtained for the Poisson GLM model
\[ E[\boldsymbol{r}] = -\frac{1}{2}(I-H)Jz, \] 
whereas the vector of variances of the Pearson residuals is given by 
\[\var[\boldsymbol{r}] = \textbf{1} + \frac{1}{2}(QHJ-T)z.\] 
For these expressions it holds that
\begin{align*}
W &= \mbox{diag}(\mu_{ij})\\   
J &= \mbox{diag}(\mu_{ij}^{1/2}) = W^{1/2} \\
Q &= \mbox{diag}(\mu_{ij}^{-1/2}) = W^{-1/2}\\
T &= 2W+I\\
H &= W^{1/2}X(X^TWX)^{-1}X^TW^{1/2} \\
Z &= X(X^TWX)^{-1}X^T,
\end{align*}
where $z$ is the diagonal of $Z$ and $\textbf{1}$ is a unit vector. 
Note that $(QHJ-T)z$ corresponds to the diagonal of $(QHJ-T)Z$ and since  
\[ QHJ =  W^{-1/2}W^{1/2}X(X^TWX)^{-1}X^TW^{1/2}W^{1/2} = X(X^TWX)^{-1}X^TW, \] 
we obtain
\begin{align*}
 (QHJ-T)Z &=  \left(X(X^TWX)^{-1}X^TW\right)\left(X(X^TWX)^{-1}X^T\right) - (2 W + I)\left(X(X^TWX)^{-1}X^T\right)\\
&=  \left(X(X^TWX)^{-1}X^T\right) - (2W + I)\left(X(X^TWX)^{-1}X^T\right)\\
&=  -2 W \left(X(X^TWX)^{-1}X^T\right) \\
&=  -2 WX \left(X^TWX\right)^{-1}X^T.
\end{align*}
Therefore we obtain \[\var[\boldsymbol{r}] = \mbox{diag}(I) - \mbox{diag}(WX \left(X^TWX\right)^{-1}X^T)\]
Since the diagonal of $W\left(X(X^TWX)^{-1}X^T\right)$ equals the 
diagonal of the hat matrix $H = W^{1/2}\left(X(X^TWX)^{-1}X^T\right)W^{1/2}$, we finally obtain that 
\begin{equation}\label{eq:PoissonGLM_Var}\var[\boldsymbol{r}] = \mbox{diag}(I-H). \end{equation}
Rewriting the expression for $E[\boldsymbol{r}]$ we obtain that 
\begin{equation}\label{eq:PoissonGLM_Exp}E[\boldsymbol{r}] = \mbox{diag}\left(-\frac{1}{2}(I-H)HW^{-1/2}\right). \end{equation}
In practice the matrices $H$ and $W$ may be replaced by their sample estimates.

\subsection{Calculations for Fast and Robust bootstrap for GLM}
\label{sec:FRB}
Since the term $E \left[\psi_c(r_{ij})\right]$ is not differentiable using its 
exact expression, we calculate this term by approximating the Poisson 
distribution with a $\chi^2$ distribution and then using the Wilson-Hilferty approximation, see \cite[chapter 4, p.~162]{Johnson:UniDisCist}. This leads to the following lemma:

\begin{lemma}
\label{Pois:Approx}
If $Y_{ij}\sim\mbox{Poisson}(\mu_{ij})$ it holds that
\[\Prob{Y_{ij} \leqslant y} \approx \left( 2\pi \right)^{-1/2} \int_{z}^{\infty}{\exp\left(-\frac{u^2}{2}\right) \, du}\]
where 
\[z = 3\left[ \left( \frac{\mu_{ij}}{y+1} \right)^{1/3} - 1 + \frac{1}{9(y+1)} \right](y+1)^{1/2}. \]
\end{lemma}

Using Lemma (\ref{Pois:Approx}) we may rewrite equation (\ref{eq:Cantoni_ExpHuberRes}) as follows: 
\begin{align*}
E \left[\psi_c\left( \frac{y_{ij} - \mu_{ij}}{\mu_{ij}^{1/2}} \right) \right] \approx \, & 
c\left[1 - \left( 2\pi \right)^{-1/2} \left(
\int_{\widehat{z}_2}^{\infty}{\exp(-\frac{u^2}{2}) \, du} + 
\int_{\widehat{z}_1}^{\infty}{\exp(-\frac{u^2}{2}) \, du}\right) \right]\\
&+(2\pi)^{(-1/2)}\mu_{ij}^{1/2}\left[ - \int_{\widetilde{z}_1}^{\widehat{z}_1}{\exp(-\frac{u^2}{2}) \, du} + 
\int_{\widetilde{z}_2}^{\widehat{z}_2}{\exp(-\frac{u^2}{2}) \, du} \right]\\
\end{align*}
with
\[\widetilde{z}_l = 3\left[ \left(\frac{\mu_{ij}}{j_l+0.5}\right)^{1/3} - 1 + \frac{1}{9(j_l+0.5)} \right](j_l+0.5)^{1/2},\]
\[\widehat{z}_l = 3\left[ \left(\frac{\mu_{ij}}{j_l+1.5}\right)^{1/3} - 1 + \frac{1}{9(j_l+1.5)} \right](j_l+1.5)^{1/2},\]
and $j_1 = \mu_{ij} - c\mu_{ij}^{1/2}$ and $j_2 = \mu_{ij} + c\mu_{ij}^{1/2}$. Note that 
in turn $j_1$ and $j_2$ are approximated by considering their non-truncated versions. However $j_1$ must be bounded from below by zero to ensure all square roots may be calculated without problems. 
Clearly we need to calculate the terms 
\[ \frac{\partial}{\partial\thetab} \int_{\widehat{z}_i}^{\infty}{\exp(-\frac{u^2}{2}) \, du} \text{ and } \frac{\partial}{\partial\thetab} \int_{\widetilde{z}_i}^{\widehat{z}_i}{\exp(-\frac{u^2}{2}) \, du}. \]
To calculate the first term note that \[\frac{\partial}{\partial\thetab} \int_{\widehat{z}_i}^{\infty}{\exp(-\frac{u^2}{2}) \, du} = -\frac{\partial}{\partial\thetab} \int_{0}^{\widehat{z}_i}{\exp(-\frac{u^2}{2}) \, du} = - \exp(-\frac{\widehat{z}_i^2}{2}) \frac{\partial}{\partial\thetab}\widehat{z}_i \]
where we used the generalized Leibniz integration rule.
Moreover,
\[ \frac{\partial}{\partial\thetab} \int_{\widetilde{z}_i}^{\widehat{z}_i}{\exp(-\frac{u^2}{2}) \, du} = \exp(-\frac{(\widehat{z}_i)^2}{2})\frac{\partial}{\partial\thetab}\widehat{z}_i - \exp(-\frac{(\widetilde{z}_i)^2}{2})\frac{\partial}{\partial\thetab}\widetilde{z}_i.\]

This leads to 
\begin{align*}
\frac{\partial}{\partial\thetab}E \left[\psi_c( \frac{y_{ij} - \mu_{ij}}{\mu_{ij}^{1/2}} ) \right] \approx \, & 
c\left[\left( 2\pi \right)^{-1/2} \left(
\exp(-\frac{\widehat{z}_2^2}{2}) \frac{\partial}{\partial\thetab}\widehat{z}_2 + 
\exp(-\frac{\widehat{z}_1^2}{2}) \frac{\partial}{\partial\thetab}\widehat{z}_1\right) \right]\\
&+\frac12 (2\pi)^{(-1/2)} \mu_{ij}^{1/2}x_i \left[ \int_{\widetilde{z}_2}^{\widehat{z}_2}{\exp(-\frac{u^2}{2}) \, du} - \int_{\widetilde{z}_1}^{\widehat{z}_1}{\exp(-\frac{u^2}{2}) \, du}\right]\\
&+ (2\pi)^{(-1/2)}\mu_{ij}^{1/2} \left[ 
\exp(-\frac{(\widehat{z}_2)^2}{2})\frac{\partial}{\partial\thetab}\widehat{z}_2 - \exp(-\frac{(\widetilde{z}_2)^2}{2})\frac{\partial}{\partial\thetab}\widetilde{z}_2 \right.\\
&- \left.
\exp(-\frac{(\widehat{z}_1)^2}{2})\frac{\partial}{\partial\thetab}\widehat{z}_1 + \exp(-\frac{(\widetilde{z}_1)^2}{2})\frac{\partial}{\partial\thetab}\widetilde{z}_1
 \right]. \\
\end{align*}
Differentiating $\widetilde{z}_l$ and $\widehat{z}_l$ yields 
\begin{align*}
\frac{\partial}{\partial\thetab}\widetilde{z}_l =& 3\left[ \frac{\partial}{\partial\thetab}\left( \frac{\mu_{ij}}{j_l+0.5} \right)^{1/3} + \frac{\partial}{\partial\thetab}\frac{1}{9(j_l+0.5)} \right](j_l+0.5)^{1/2}\\
&+3\left[ \left( \frac{\mu_{ij}}{j_l+0.5} \right)^{1/3} - 1 + \frac{1}{9(j_l+0.5)} \right]\frac{\partial}{\partial\thetab}\left((j_l+0.5)^{1/2}\right),\\
\frac{\partial}{\partial\thetab}\widehat{z}_l =& 3\left[ \frac{\partial}{\partial\thetab}\left( \frac{\mu_{ij}}{j_l+1.5} \right)^{1/3} + \frac{\partial}{\partial\thetab}\frac{1}{9(j_l+1.5)} \right](j_l+1.5)^{1/2}\\
&+3\left[ \left( \frac{\mu_{ij}}{j_l+1.5} \right)^{1/3} - 1 + \frac{1}{9(j_l+1.5)} \right]\frac{\partial}{\partial\thetab}\left((j_l+1.5)^{1/2}\right).
\end{align*}
Moreover, 
\begin{align*}
\frac{\partial}{\partial\thetab}\left( \frac{\mu_{ij}}{j_l+m} \right)^{1/3} &= \frac13 \frac{\mu_{ij}^{1/3}}{(j_l+m)^{4/3}} \left(x_{ij}(j_l+m) - \frac{\partial}{\partial\thetab} j_l \right), \\
\frac{\partial}{\partial\thetab}\frac{1}{9(j_l+m)} &= \frac{-\frac{\partial}{\partial\thetab} j_l}{9(j_l+m)^2},\\
\frac{\partial}{\partial\thetab}\left((j_l+m)^{1/2}\right) &= \frac12 \frac{1}{(j_l+m)^{1/2}} \frac{\partial}{\partial\thetab} j_l,
\end{align*}
with $\frac{\partial}{\partial\thetab} j_1 = x_{ij}(\mu_{ij}-\frac{c}{2}\mu_{ij}^{1/2})$ and 
$\frac{\partial}{\partial\thetab} j_2 = x_{ij}(\mu_{ij}+\frac{c}{2}\mu_{ij}^{1/2})$.

\subsection{Proof of theorem \ref{theo:SIF}}
\label{sec:ProofOfTheorem}
The estimator by \cite{Cantoni:genreg} may be generally written as 
\[ \sum_{i = 1}^{n}\sum_{j=1}^{n-i+1}{\psi\left(y_{ij},\mu_{ij}\right) } = 0, \]
where \[\psi\left(y_{ij},\mu_{ij}\right) = \nu\left(y_{ij},\mu_{ij}\right)w\left(x_{ij}\right)\mu_{ij}'-a\left(\boldsymbol{\thetab}\right)\] can be seen as the score function and 
$a\left(\boldsymbol{\thetab}\right) = \frac1N \sum_{i = 1}^{n}\sum_{j=1}^{n-i+1}E\left[\nu\left(y_{ij},\mu_{ij}\right)\right]w\left(x_{ij}\right)\mu_{ij}'$. 
This is the structure of an $M$-estimator which leads to the following influence function: 
\[ \IF\left([z,l,m], T_{\thetab}, F\right) = M\left(\psi, F\right)^{-1}\psi\left(z,\mu_{lm}\right),\] 
with $M\left(\psi, F\right) = -E\left[\frac{\partial}{\partial\thetab}\psi\left(z,\mu_{lm}\right)\right]$. 
More details may be found in e.g. \cite{Hampel:IFapproach}. The asymptotic variance can then be found as \[V(\psi,F) = M\left(\psi, F\right)^{-1}Q\left(\psi, F\right)M\left(\psi, F\right)^{-t}, \] 
where $Q\left(\psi, F\right) = E\left[\psi\left(z,\mu_{lm}\right)\psi\left(z,\mu_{lm}\right)^t\right]$.
Using the above expressions, the standardised influence function of the $M$-estimator becomes
\[ \left( \IF^t \, V^{-1} \, \IF \right)^{\frac12} = \left(\psi^t M^{-t} M^t Q^{-1} M M^{-1} \psi\right)^{\frac12}  = \left(\psi^t Q^{-1} \psi\right)^{\frac12}, \]
where the arguments of the function have been dropped for lightning the notation. 

Choosing $\nu\left(z,\mu_{lm}\right) = \psi_c(r_{lm}) \frac{1}{V^{1/2}(\mu_{lm})}$ leads to the robust estimator of \cite{Cantoni:genreg} as detailed in section \ref{sec:RobustGLMEstimator}. However, for the influence function bootstrap procedure we need the expression for the influence function of the classical estimator. Therefore, we choose $\nu\left(z,\mu_{lm}\right) = \frac{r_{lm}}{V^{1/2}(\mu_{lm})}$ and set $w\left(\cdot\right) = 1$ which are the functions corresponding to the Poisson GLM yielding the same estimates as the classical chain-ladder. Thus \eqref{eq:RESIFDef} reduces to 

\begin{equation}
\label{eq:ProofIFBA}
\mbox{RESIF}([z,l,m],T_{\thetab}^{nr},F_{\hat{\thetab}^{r}}) = \left[ \frac{\psi\left(z,\hat{\mu}_{lm}\right)^t\psi\left(z,\hat{\mu}_{lm}\right)}{Q\left(\psi, F_{\hat{\thetab}^{r}}\right)}  \right]^{\frac{1}{2}}. 
\end{equation}

It is easily checked that $E[r_{lm}] = 0$, which leads to $a\left(\thetab\right)=0$. Hence, $\psi\left(z,\hat{\mu}_{lm}\right) = r_{lm}\hat{\mu}^{1/2}_{lm}$ and thus \eqref{eq:ProofIFBA} becomes 
\[ \mbox{RESIF}([z,l,m],T_{\thetab}^{nr},F_{\hat{\thetab}^{r}}) = \frac{|r_{lm}|\mu^{1/2}_{lm}}{\left(E\left[r_{lm}^2\mu_{lm}\right]\right)^{1/2}}. \] This immediately simplifies to 
\[ \mbox{RESIF}([z,l,m],T_{\thetab}^{nr},F_{\hat{\thetab}^{r}}) = \frac{|z-\mu_{lm}|}{\mu_{lm}^{1/2}} \] 
which proves the theorem.



\end{document}